# Is there a genetic relationship between chondrules and matrix?

Elishevah M.M.E. van Kooten[1*], Adrian Brearley[2], Denton S. Ebel[3,4,5], Conel M.O.'D. Alexander[6], Marina E. Gemma[7,3], Dominik C. Hezel[8]

[1]Centre for Star and Planet Formation, Globe Institute, University of Copenhagen, Øster Voldgade 5-7, 1350 Copenhagen, Denmark,
[2]Department of Earth & Planetary Sciences, University of New Mexico, Albuquerque, New Mexico 87131, USA,
[3]Department of Earth and Planetary Sciences, American Museum of Natural History, New York, NY 10024, USA;
[4]Department of Earth and Environmental Sciences, Columbia University, New York;
[5]Graduate Center of City University of New York, NY, USA;
[6]Carnegie Institution of Washington, 5241 Broad Branch Road, Washington DC, 20015, USA,
[7]Department of Geosciences, Stony Brook University, Stony Brook, NY, 11974, USA;
[8]Goethe-Universität Frankfurt, Altenhöferallee 1, 60438 Frankfurt am Main, Germany

*Corresponding author: elishevah.vankooten@sund.ku.dk

**Abstract**

Chondritic components such as chondrules and matrix are the key time capsules that can help us understand the evolution and dynamics of the protoplanetary disk from which the Solar System originated. Knowledge of where and how these components formed and to what extent they were transported in the gaseous disk provides major constraints to astrophysical models that investigate planet formation. Here, we explore whether chondrules and matrix are genetically related to each other and formed from single reservoirs per chondrite group or if every chondrite represents a unique proportion of components transported from a small number of formation reservoirs in the disk. These 'static versus dynamic disk' interpretations of cosmochemical data have profound implications for the accretion history of the planets in the Solar System.

To fully understand the relationship between chondrules and matrix and their potential "complementarity", we dive into the petrological nature and origin of matrix, the chemical and isotopic compositions of chondrules and matrix and evaluate these data considering the effect of secondary alteration observed in chondrites and the potential complexity of chondrule formation. Even though we, the authors, have used different datasets and arrived at differing interpretations of chondrule-matrix relationships in the past, this review provides clarity on the existing data and has given us new directions towards future research that can resolve the complementarity debate.

## 1. Introduction: What is complementarity and (why) is there a problem?

Chondrites are aggregates of components that formed in the protoplanetary disk surrounding the young Sun. The major chondrite components are chondrules, matrix, and to a lesser extent, refractory inclusions (CAIs: calcium-, aluminum-rich inclusions, and AOAs: amoeboid-olivine-aggregates), and opaque phases such as metal, magnetite or sulfide (in rare cases up to 70 vol.% metal). A key cosmochemical constraint on the formation, evolution and dynamics of the protoplanetary disk is whether these individual components formed together in local disk regions, or whether they formed in separate regions of the protoplanetary disk and were later transported, mixed, and accreted into chondrite parent bodies. For example, it is crucial to understand if terrestrial planets accreted only from local feeding zones or after significant radial mass transport of dust and pebbles (**Fig. 1**). Moreover, the debate over whether there is a genetic relationship between chondrules and matrix (i.e., complementarity) has far-reaching astrophysical implications for how disks function and planetesimals form. Chondrules and matrix generally represent the main mass of chondrites, so we focus on understanding the potential genetic relationships between these two components, although we do not disregard the potential contributions of refractory inclusions to the chondrule (and matrix) forming regions.

*Chemical* complementarity is defined as opposing deviations of element ratios in chondrules and matrix of a chondrite with a CI bulk chondrite value of the considered element. We assume a close to CI chondrite chemical composition of the initial dust inventory of the protoplanetary disk, since CI chondrites most closely match the solar photosphere for the non-volatile elements (Lodders et al., 2021; Barrat et al., 2012; Frank et al., 2023). Thermal processing of CI-like dust may have altered its volatile composition, but as we restrict ourselves to the moderately volatile and more refractory elements, this has no bearing on our discussion. As the reader will notice throughout the chapter, this definition for elemental ratios is not applicable to *isotopic* complementarity since bulk chondrites are generally isotopically not CI-like due to inherited heterogeneities from the molecular cloud and/or processing of dust in the disk. Moreover, since non-carbonaceous chondrites typically have minor fractions of matrix and non-solar bulk ratios for many elements, we only discuss carbonaceous chondrites in this chapter. The complementarity model assumes that chondrules and matrix initially formed from a dust reservoir with a solar (CI-like) composition and, respectively, became sub- and supersolar during chondrule formation events (Palme et al., 2015). Complementarity requires that chondrules formed in close proximity to the matrix that envelopes them and that despite very large differences in grain size the two remained coupled until accretion into their parent bodies.

The interpretation of observed complementary chondrule-matrix compositions largely hinges on whether the deviations in the average compositions of chondrules and matrix from the solar composition can be explained by exchange of materials prior to, during or after chondrule formation. We approach the debate by first defining the nature and origin of the matrix and chondrules, as well as discussing the existing and controversial elemental and isotopic datasets. One of the aims of this study is trying to understand whether these datasets are consistent with one another and to assess the pros and cons of the various approaches adopted to date. Following this, we will test these datasets in the framework of secondary alteration in the chondrite parent body and/or from terrestrial weathering. We then place these data in the context of astrophysical models of protoplanetary disk dynamics. Finally, we aim to construct several lines of future research that should be followed to ultimately create a paradigm for the relationship between chondrules and matrix.

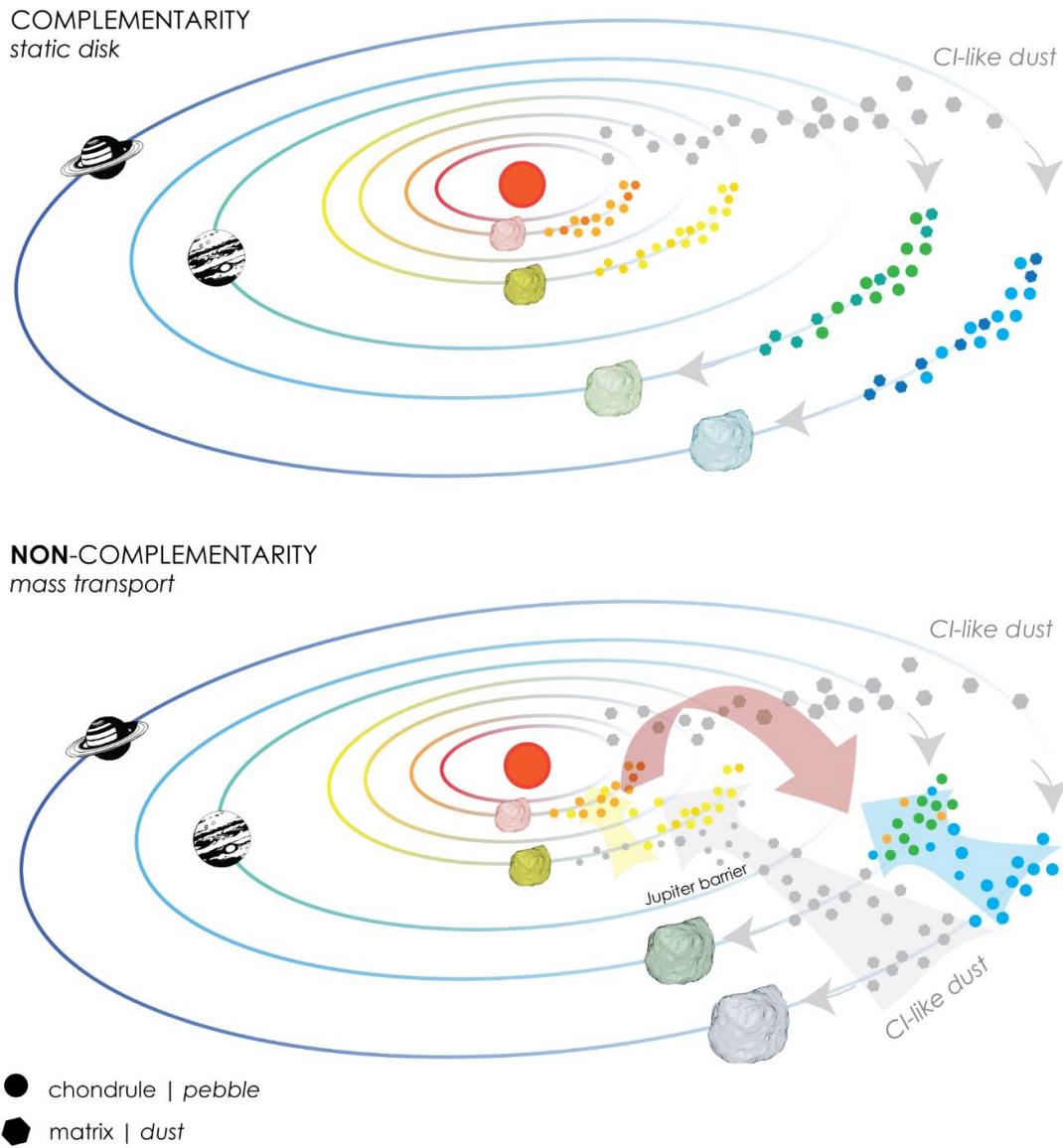

*Fig. 1: A schematic representation of the early Solar System, with two opposing models of chondrule (pebble) and matrix (dust) mass transport. A) Matrix and chondrules form co-genetically from CI-like dust and accrete together in local reservoirs. Mixing between reservoirs is limited and matrix and chondrules are complementary to each other. B) Transport of chondrules by mass outflows and inward flux of CI-like dust combined with 'pebble traps' (Whipple, 1972) result in non-complementary relationships between chondrules and matrix.*

## 2. What is the petrological nature of matrix and chondrules?

### 2.1 Definition of matrix

Any consideration of the question of compositional complementarity between chondrules and matrix needs to consider the exact definition of matrix. Matrix has traditionally been defined as the dark, fine-grained material that occurs interstitially to chondrules and essentially represents the 'glue' which holds chondrules together to form a rock. Scott et al. (1988) defined matrix as material which has a grain size of <5 micron. This has become a widely used working definition of matrix that has stood the test of time but is worthy of re-evaluation based on more recent observations. Firstly, using the term 'dark' to define the matrix is problematic, because it simply describes its appearance in transmitted light in standard (30 μm thick) petrographic thin sections, not an inherent characteristic of the matrix. Although the matrix is the host of most organic matter in chondrites, the concentrations are not sufficient to make the matrix dark. Therefore, it is most appropriate to just use the term fine-grained to describe matrix, especially since many modern studies do not involve thin section microscopy.

The issue of grain size is also important. Studies of the most pristine chondrites show that matrix has two major types of occurrences: fine-grained rims surrounding chondrules and other coarse-grained objects, and as interstitial material between chondrites. In some chondrites a significant proportion of the matrix occurs as fine-grained rims, whereas in other chondrites, fine-grained rims are relatively rare. For example, Metzler et al. (1992) showed that in some rare CM chondrites that had escaped brecciation, essentially all the fine-grained matrix occurred as rims around chondrules, refractory inclusions, etc., a feature that they termed a primary accretionary texture. With increasing brecciation, the abundance of fine-grained rims decreased due to regolith comminution and the fine-grained component of the chondrites became dominated by a clastic matrix that consisted of chondrule fragments, mineral grains derived from chondrules, and material derived from fine-grained rims. Most CM chondrites are breccias (Metzler et al. 1992; Lentfort et al. 2021; Bischoff et al., 2017) and therefore matrix in these meteorites is a complex mixture of fine-grained matrix and material derived from chondrules and other coarse-grained objects. On the other hand, in CO chondrites such as ALH 77307, many chondrules have fine-grained rims (**Fig. 2A**), but there is also a primary interstitial matrix component that does not occur as fine-grained rims (e.g., Brearley, 1993). In general, CO3 chondrites have escaped the extensive brecciation that has affected the CM chondrites. In the CR chondrites, fine-grained rims are rare and most of the matrix is interstitial to chondrules and other coarse-grained objects (e.g., Weisberg et al. 1993; Abreu and Brearley, 2010; Schrader et al. 2011). Although CR chondrites are clearly breccias (Bischoff et al. 1993; Weisberg et al. 1993), the extensive comminution that affected most CM chondrites has not occurred in the CRs and their matrices consist largely of extremely fine-grained material with minor amounts of chondrule fragments.

This work provides a framework for reevaluating what we define as matrix and hence the grain size (diameter) of what constitutes matrix material. It has become increasingly apparent that in the most pristine chondrites that have escaped significant brecciation, mineral grains with sizes >1 micron are rare (**Fig. 2B and C**). In addition, electron probe microanalytical (EPMA) studies of olivines and pyroxenes from the matrices of a range of type 3 ordinary and carbonaceous chondrites (i.e., grain sizes >3 microns) (Matsunami, 1984; Alexander et al. 1989b; Frank and Zolensky, 2014; Patzek et al 2021) show that essentially all these grains have major and minor element compositions that were consistent with derivation from either type I or type II chondrules. This suggests then that the coarser-grained material (i.e., greater than ~2 microns in size) that may occur embedded within matrix should not be considered to be a part of matrix but originates from chondrules that presumably underwent fragmentation in the solar nebula and were subsequently accreted with chondrules. Additional fragmentation likely occurred as a result of regolith processes during accretion. These observations lead to the conclusion that a better definition of matrix in terms of grain size, especially when considering the question of complementarity, is that it consists almost exclusively of material that is submicron in character. This material most likely represents the component of nebular dust that must be understood fully to constrain whether matrix and chondrules have complementary compositions, a task which is by no means straightforward.

## 2.2 The nature of matrix

In considering the question of whether chondrules and matrix are complementary, it is important to have a complete understanding of the petrologic, chemical, and isotopic characteristics of both types of components. Although chondrules have been studied in extensive detail and their characteristics are generally well understood in the different chondrite groups, the same cannot be said of matrix, although there have been important advances over the last 20 years. There are a number of reasons why this is the case. First, although it seems to be relatively straightforward to obtain the elemental composition of matrix, it turns out that this is not necessarily straightforward for both analytical and sample reasons, as discussed later (section 3). Second, determining the mineralogical characteristics of matrix is hampered by its fine-grained nature that requires highly specialized instrumentation such as transmission electron microscopy (TEM) that is not widely available. Third, there have been significant misconceptions about matrix that arise from a lack of understanding of the effects of thermal metamorphism and aqueous alteration on the characteristics of matrix (section 5). Due to its fine-grained characteristics, matrix is highly sensitive to both heating and interaction with aqueous fluids. Although petrologic type 3 chondrites are considered to be the least altered of all chondrites, within the type 3s, it is only the very lowest type 3s, with petrologic type <3.1, that come close to preserving the pre-accretionary characteristics of matrix. For example, Grossman and Brearley (2005) demonstrated that there are important and detectable changes in the Cr content of ferroan olivine in type II chondrules in UOCs and CO chondrites that occur below petrological grade 3.1. Matrix shows significant compositional variations over this range of petrologic types, notably a decrease in the S content due to rapid recrystallization of very fine-grained sulfides into coarser grains. A number of studies (e.g., Alexander et al. 1989a,b; Brearley, 1993; Brearley and Jones, 1998; Dobrică and Brearley, 2020a; Brearley, 1993; Krot et al., 1995, 1997) demonstrate that by petrologic type 3.1, the matrix is essentially completely altered and has lost all of its primary textural and mineralogic characteristics. The compositional changes that may accompany these mineralogical changes are incompletely understood at present (see section 5).

Despite these caveats, there is a significant body of evidence that at least some chondrites do retain important information about the characteristics of matrix which likely do reflect the primary characteristics of nebular dust. Over the past three decades, a small number of chondrites in several of the carbonaceous chondrite groups and the ordinary chondrites have been recognized to show minimal degrees of thermal metamorphism. Recognizing these specific chondrites has become possible through a combination of different kinds of analyses. For instance, the Cr in olivine technique (Grossman and Brearley, 2005) has provided a straightforward method for identifying very low petrologic type chondrites. In many cases, the pristine nature of these chondrites has been confirmed by subsequent studies of isotopically anomalous presolar silicate grain abundances (e.g., Nguyen et al. 2004; Floss and Stadermann, 2009; Bose et al., 2012; Nittler et al. 2019). High abundances of presolar silicates have become a definitive signature of relatively pristine chondrites, because of these silicates' high susceptibility to destruction by thermal metamorphism and aqueous alteration (e.g., Floss and Haenecour, 2016).

There are now a handful of very low petrologic type 3 chondrites that provide important insights into matrix that has not been extensively modified by thermal metamorphism. These include the anomalous, but heavily weathered, C chondrite Acfer 094 (3.00), the CO3 chondrites DOM 08006 (3.00), and ALHA 77307 (3.01), and regions of matrix in the LL3.00 chondrite Semarkona (Greshake, 1997; Davidson et al. 2019; Brearley, 1993; Dobrică and Brearley, 2020a). Additionally, there are now a number of meteorites that are considered to be petrologic type 2 chondrites that exhibit minimal evidence of aqueous alteration and also preserve matrices that are comparatively unaltered, such as the CM chondrites Paris (Hewins et al. 2014; Leroux et al. 2015), Asuka 12169, and Asuka 12236 (Kimura et al. 2020; Noguchi et al. 2021), as well as the CR chondrites QUE 99177, MET 00426, and GRV 021710 (Abreu and Brearley, 2010; Le Guillou and Brearley, 2014; Davidson et al. 2019). It is, however, important to recognize that almost all CM chondrites are breccias and consist of clasts that have experienced variable degrees of alteration, as documented in Paris (Hewins et al. 2014). Moreover, the Antarctic finds as well as Acfer 094 have undergone weathering to varying degrees (Jull et al., 1988; Velbel, 1988; Bland et al., 2005; Tyra et al., 2007).

The sample suite of minimally altered chondritic meteorites is, therefore, still very limited and few studies have used these samples to address the question of chemical complementarity (Ebel et al., 2016; Patzer et al., 2021, 2022, 2023). However, these meteorites have provided a much clearer picture of the constituents of matrix that provide a basis for evaluating if and how matrix is related to chondrules. Observations of these samples also provide a much more coherent view of matrix across different chondrite groups that was previously obscured by the fact that studies of matrix were carried out on type 3 chondrites with varying petrologic types. As a result, the pervasive view of matrix until the early 1990s was that it consisted predominantly of ferroan olivine that was the product of nebular condensation (e.g., Nagahara, 1984; Palme and Fegley, 1990), although a wide range of other mechanism had also been proposed (e.g., Housley and Cirlin, 1983; Kojima and Tomeoka, 1995; Krot et al. 1997). This conclusion has now been demonstrated to be incorrect. The ferroan olivine in matrices of type 3 carbonaceous and ordinary chondrites is principally the product of metamorphic recrystallization of amorphous material (e.g., Alexander et al., 1989a; Dobrică and Brearley, 2020b).

## 2.3 Characteristics of matrix in the most pristine chondritic meteorites

Although there is wide disagreement on the exact formation mechanism of chondrules, there is general agreement that chondrules were all likely formed by similar heating mechanisms in most chondrite groups (with the exception of the CB and CH chondrites; Krot et al., 2005). Until recently, this type of relatively coherent view was lacking for matrix. Indeed, a different model existed for the formation of matrix in every chondrite group (e.g., Scott et al., 1988). The lack of any consistent formation mechanism that might be relevant to matrix in all chondrite groups arose from the study of matrix in chondrites that had experienced variable degrees of thermal metamorphism, even if in some cases this was relatively low based on apparent petrologic type. But as noted above, even type 3.1 chondrites do not contain pristine matrix (Brearley and Jones, 1998).

This situation has changed significantly as a result of studies of the low petrologic type 3 chondrites and minimally altered type 2 chondrites. It is now apparent that matrix in all the chondrite groups shares a remarkable number of similarities, suggesting that like chondrules, matrix is likely the product of similar processes in all the chondrite groups. This is a very important step forward in understanding matrix and the relationship of matrix to chondrules. These characteristics will be discussed below.

Our concept of matrix is slowly becoming more sophisticated. This is also extremely important for understanding the question of complementarity. Most studies have viewed matrix as some kind of homogeneous component that consists of just one kind of material. However, the recognition that matrix consists of different kinds of materials that are the result of different kinds of processes that may have occurred at different times and locations within the protoplanetary disk is slowly emerging. Alexander (2005) was among the first to point out that matrix should be considered as consisting of a mixture of different components with potentially different origins, consistent with the observation that matrix consists of a highly unequilibrated mixture of different materials (e.g., Brearley, 1993). Indeed, it is reasonable to propose that matrix should actually be viewed almost as we view chondrites on a macroscopic scale, i.e., consisting of different components, such as CAIs, chondrules, Fe,Ni metal etc., but on the submicron scale. The distinct components that represent constituents of matrix are discussed below and are based on analyses carried out on both macroscopic samples of matrix and by microbeam and nanobeam techniques such as nanoSIMS and TEM.

## 2.4 Major components of matrix

*Amorphous silicates:* Alexander et al. (1989b) first reported the presence of amorphous silicates in the matrices of UOCs. The first occurrence in a type 3 carbonaceous chondrite was described by Brearley (1993) in the CO3 chondrite ALHA 77307. Subsequent studies have demonstrated that the widespread occurrence of amorphous silicates is a characteristic of the matrices of Acfer 094 (**Fig. 2E**; Greshake, 1997; Ohtaki et al. 2021), as well as the CM and CR chondrites noted above and pristine regions of the matrix of Semarkona (Dobrică and Brearley, 2020a). This material is widespread and is consistently iron-rich. In some chondrites, the amorphous silicates appear to occur as distinct domains that might represent discrete, primary nebular dust grains (e.g., Yamato 791198 CM2; Chizmadia and Brearley, 2008). These distinct domains are defined by the presence or absence of sulfide grains. However, in other chondrites, such as the CO and CR chondrites, these domains are not so readily apparent. Nevertheless, at the submicron scale, there are clearly significant heterogeneities (**Fig. 2F**). These amorphous silicates have been suggested to have affinities to GEMS (glass with embedded metal sulfide) that occur in IDPs (Bradley, 1994; Leroux et al. 2015), although in detail, there are important differences between the two types of materials. As discussed in detail by Le Guillou and Brearley (2014), Le Guillou et al. (2015) and Ohtaki et al. (2021), the amorphous silicates in even the least altered chondrites have undergone hydration and oxidation, and consequently have been extensively modified since their formation. The absence of metal grains in these GEMS-like materials within chondrite matrices and the high Fe contents of the amorphous silicates further distinguishes them from GEMS in IDPs (Ohtaki et al. 2021).

*Nanophase Fe-sulfides.* Another ubiquitous feature of the most pristine chondrites is the occurrence of nanophase Fe and Fe-, Ni-sulfides, principally pyrrhotite and pentlandite (**Figs. 2D and F**). These phases occur embedded with the Fe-rich amorphous silicates and can have variable distributions and abundance at the submicron scale. In some chondrites, distinct boundaries exist between sulfides embedded in GEMS-like materials and sulfide-free domains containing, for example, amorphous silicates and organics. In other chondrites, this distinction between sulfide and sulfide-rich domains is not as well defined.

*Crystalline anhydrous silicates.* A common misconception from early studies of matrix was that it was dominated by Fe-rich olivine. This is now known to be incorrect. The abundance of crystalline anhydrous silicates is variable but generally low in pristine chondrite matrices. The highest abundance is in Acfer 094 with ~30 vol.%. In comparison the CR, CO and CM chondrites contain typically <15 vol.%. However, more crystal-rich regions of matrix have been observed, such as in fine-grained rims in some CM chondrites (e.g., QUE 97990; Brearley et al., 2016). The distribution and characteristics of anhydrous silicates is variable, but they are dominated by Mg-rich olivines and pyroxenes, the latter occurring in both ortho and clinoenstatite forms, and sometimes as disordered intergrowths of the two. These anhydrous silicates occur as isolated grains or aggregates, some of which have textures suggesting that they have been annealed to some degree but not to the extent that they have undergone significant grain growth. Grain sizes of individual grains and aggregates are rarely larger than a micron in size and are more typically 0.3 to 0.7 microns. Although compositional data are still relatively sparse, all chondrite groups contain matrix olivine and pyroxenes which are predominantly Mg-rich. There is only rare evidence of Fe-bearing olivines and pyroxenes. Remarkably, evidence of anhydrous minerals such as those that are constituents of CAIs is extremely rare although some evidence for microCAI-like objects has been reported in Acfer 094 (Bland et al. 2007). The apparent lack of refractory minerals such as hibonite, melilite, perovskite and spinel, for example, may be a combination of the fact that the volume of matrix that has been studied at the TEM scale is extremely low and that these minerals may be heterogeneously distributed.

*Presolar grains.* It is now fully recognized that the fine-grained matrices of chondrites are the hosts of the presolar grain populations in chondrites, including diamonds, silicon carbide, graphite, TiN, etc., as well as presolar silicate grains (Zinner, 2014; Hoppe et al., 2022; Nittler and Ciesla, 2016). NanoSIMS studies of matrices now routinely identify these grains in situ in matrix in carbonaceous and ordinary chondrites using isotopic mapping techniques (e.g., Nguyen et al 2004, 2007; Floss and Staderman, 2009; Bose et al. 2012; Nittler et al. 2021) The highest abundances are present in the most pristine carbonaceous chondrites (as noted above) and also in unequilibrated ordinary chondrites such as Semarkona (Barosch et al. 2022). These grains represent a distinct component of matrix, which, although present at low abundances (10s to 100s of ppm in the most pristine chondrites, e.g., Floss and Haenecour, 2016) signifies the presence of a presolar component that escaped thermal processing within the protoplanetary disk. The abundances of the major types of presolar grains are fairly uniform amongst the most primitive chondrite matrices (e.g., Huss and Lewis 1995; Davidson et al. 2014; Nittler et al. 2021), except presolar silicates that have been largely destroyed in CIs.

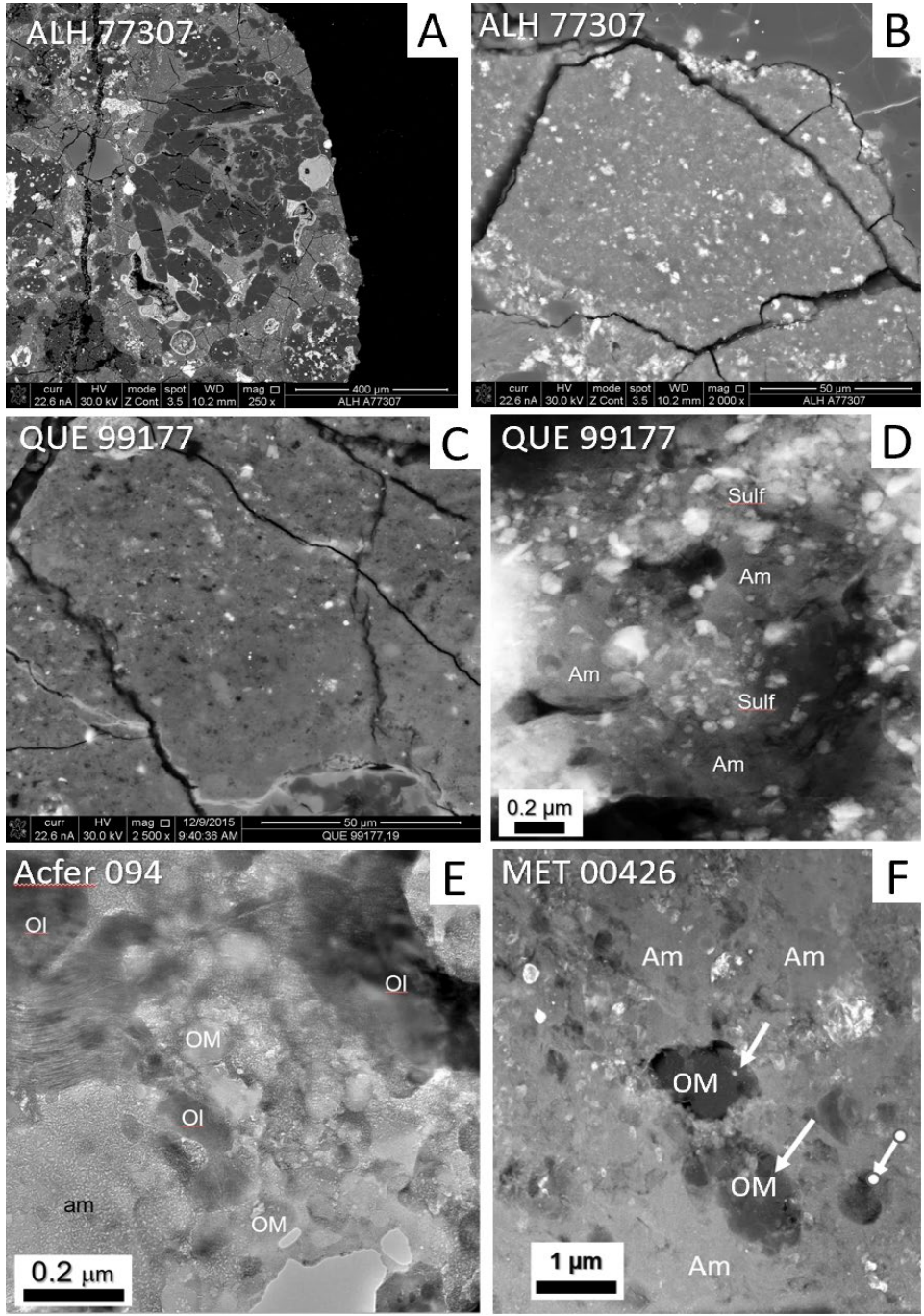

*Figure 2: Backscattered electron SEM, STEM and TEM images showing the characteristics of matrix in different pristine carbonaceous chondrites. A) BSE image of well-defined fine-grained rim around a porphyritic type IAB chondrule in the ALH 77307 CO3 chondrite. B) High magnification BSE image of a fine-grained rim in ALH 77307 showing the extremely fine-grained characteristics of the matrix. The bright grains, the largest component in the matrix, are magnetite grains produced by secondary alteration. The rest of the matrix consists of an extremely fine-grained, heterogeneous groundmass. Coarser-grained silicate minerals even a few microns in size such as olivine and pyroxene are extremely rare. C) BSE image of a typical region of matrix in the QUE 99177 CR chondrite showing the heterogeneous characteristics of the matrix at a fine scale. Chondrule fragments or isolated olivines and pyroxenes are present locally (lower Z phases within the matrix) but are typical <3 microns in size. D) Dark-field STEM image of region of matrix in the QUE 99177 CR chondrite that contains a high abundance of nanosulfides (sulf - grains with high Z contrast) embedded in a groundmass of amorphous Fe-rich silicate material (Am). Crystalline silicate minerals are absent in this field of view. E) Bright-field TEM image of matrix in the unique carbonaceous chondrite, Acfer 094 showing the much higher abundance of very fine-grained crystalline silicates, mainly olivine (OL) in this case, although pyroxenes are also commonly present. The olivines are embedded in a matrix of amorphous Fe-rich silicate material (am). F) Dark-field STEM image of matrix in the CR chondrite MET 00426 showing the presence of common organic matter grains (OM and white arrows, low Z contrast) embedded within amorphous Fe-rich silicate. In this region of the matrix the abundance of nanosulfides (high Z grains) is much lower than in the region of matrix of QUE 99177 shown in panel D.*

*Organic matter.* Numerous studies have shown that a wide diversity of organic matter is present in chondritic meteorites, especially the carbonaceous chondrites (e.g., Cronin and Pizzarello, 1988; Alexander et al., 2017). Two distinct types of organic matter have been identified, insoluble and soluble in acids, the former being the most abundant and representing ~70 % of the total organic material (Alexander et al., 2017). Recent studies have shown convincingly that the insoluble organic matter is predominantly located within the fine-grained matrices of chondrites (e.g., Garvie and Buseck, 2007; Zega et al., 2010; Le Guillou et al., 2014; Le Guillou and Brearley, 2014; Changela et al., 2018) and occurs with a variety of morphologies and grain sizes ranging from the micron down to the nanometer scale. The finest grained organic matter occurs intimately intergrown with inorganic components of matrix, including amorphous material, phyllosilicates, sulfides, oxides, carbides, etc. In situ nanoSIMS location and analysis of insoluble organic matter grains in pristine chondritic meteorites has shown that a significant number of grains show isotopic anomalies for H (D/H), N ($^{15}N/^{14}N$), and C ($^{13}C/^{12}C$) (e.g., Busemann et al. 2006; De Gregorio et al., 2010; Floss et al., 2014; Alexander et al., 2017). These isotopic anomalies for different elements show no specific correlations but suggest different formation locations and origins for different organic grains that are still poorly understood. However, the elevated D/H ratios found in organic grains in all chondrites are suggestive of presolar (e.g., Robert and Epstein, 1982; Yang and Epstein, 1983; 1984; Kerridge, 1983; Busemann et al. 2006) or outer solar system sources (Remusat et al. 2006, 2010). The widespread survival of isotopically anomalous organic matter grains in pristine chondrites that are highly sensitive to thermal processing demonstrates that this material, like presolar grains and amorphous silicates, is an additional distinct component that has escaped the relatively high temperature required to form chondrules. The total abundances of organic material normalized to matrix abundance in all primitive chondrite matrices are roughly CI-like (Alexander et al. 2007).

*Implications:* These mineralogical and isotopic data support arguments of Alexander (2004, 2019), who noted that matrix is not a single component but should be considered as a complex mixture of highly unequilibrated materials that formed by distinct processes in different astrophysical environments, including presolar and inner and outer protoplanetary disk environments. These materials were mixed in different proportions prior to accretion. The bulk of the silicate material in pristine chondrite matrices, namely the amorphous FeO-rich silicates (with and without embedded sulfides), conceivably has a genetic relationship to chondrules. They potentially represent a condensed evaporative residue produced by volatilization during chondrule formation (e.g., Wasson and Brearley, 1993; Greshake, 1993). This is plausibly supported by, for example, the CR chondrites, which are dominated by extremely sulfide-poor chondrules embedded within a very sulfide-rich matrix (Abreu and Brearley, 2010). Alternatively, an origin unrelated to chondrules has been proposed for these amorphous silicates with embedded sulfides through links to GEMS from IDPs (Bradley et al., 2022) or reflecting mixtures of interstellar matter and material made by heating in the disk such as by FU Orionis outbursts (Alexander et al., 2017). Regardless of their origin, the FeO-rich silicate component was clearly mixed with an organic and presolar grain bearing component that is unrelated to chondrule formation. An important observation here, is that the bulk chondritic matrix has roughly CI-like organic and presolar grain abundances, implying that the amorphous material was included in this component and was not added later to significantly dilute the CI-like matrix.

## 2.5 The petrological complexity of chondrules

Complementary compositions between bulk matrix and bulk chondrules have been observed that together amount to a solar bulk chondrite composition (section 3). Within the complementarity model, this implies that all chondrules within a given chondrite formed co-genetically with the matrix. If this model is correct, the observed complementarity should be reconcilable with the petrological complexity and the underlying formation mechanism of chondrules. This includes the presence of mineralogically and compositionally different chondrule types (i.e., Mg-rich, FeO-rich, Al-rich) within the same chondrite and the fact that many chondrules underwent multiple accretion and melting events (Fox and Hewins, 2005; Rubin, 2010; Rubin and Krot, 1996; Wasson, 1993) as evidenced by, for example, their igneous rims (Krot and Wasson, 1995; Rubin, 2010, 1984; Rubin and Wasson, 1987) and relict chondrule grains (e.g., Rambaldi, 1981; Nagahara, 1981; Jones, 2012; Marrocchi et al., 2018). Here, we explore how multiple melting events change the chemical composition of chondrules and whether the petrological range of chondrules can be explained by formation in a single reservoir.

### 2.5.1 Multiple melting events

Relatively small chondrules from CO and CM chondrites typically record a seemingly singular melting event, in which the chondrule shows a homogeneous texture. Even so, FeO-poor relict grains in type II chondrules are suggested fragments from a previous generation of type I chondrules (Schrader and Davidson, 2017). The presence of dusty olivine in type I chondrules shows that they contained fragments of type II chondrules in their precursors, and, hence, the chondrule represents multiple chondrule melting events. CM and CO chondrites contain relict grains in 12 % and 48 % of the type II chondrules, respectively (Schrader et al., 2015; Schrader and Davidson, 2017; Tenner et al., 2013), suggesting that a significant portion of the chondrules is derived from a non-solar starting composition. This means that either a previous chondrule melting would have produced a chondrule with a non-solar composition (i.e., by volatile or physical loss of a component) or that material with solar composition accreted to non-solar relict grains and melted to produce a secondary chondrule. Even for completely molten chondrules without relict grains, it is possible that these chondrules also experienced multiple melting events of which only evidence for the last is preserved in the chondrule record. As such, in the complementarity model, chondrules and matrix are required to experience repeated melting events together to maintain their observed complementary compositions. To keep chondrules and fine-grained dust physically together in the turbulent protoplanetary disk, as required by complementarity, means that these melting events occurred at rapid intervals. In this case, the matrix would have to have experienced multiple thermal processing events as well.

The matter becomes more complicated for larger chondrules observed in CV and CR chondrites (**Fig. 3**). These chondrules can exhibit multiple metal and silicate rim structures, with different textures (i.e., barred/radial versus porphyritic) and temperature profiles. One of the most complex examples of these onion shell structures is shown in **Figure 3B**, where three metal and four silicate rims can be observed in a single CR chondrule. This chondrule is suggested to have formed from the accretion of different metal and silicate dust populations that were flash heated in highly local events (Hobart et al., 2015). Alternatively, similar multilayered chondrules may instead represent enveloping compound chondrules produced by collisions and sectioned near the contact (Jacquet et al., 2021). Although not all chondrules in CV and CR chondrites show the same degree of complexity, a significant percentage of them reflect at least two melting events and, thus, two accretion events. Hewins (1999) noted that although chondrules clearly experienced multiple accretion events, it has so far not been elucidated at what time intervals these events occurred. Hence, constraining the relative timing of these events (preferably in combination with nucleosynthetic isotope systematics, see section 4) would significantly increase our understanding of the locality and frequency of chondrule formation and, therefore, on the likelihood of single reservoir formation of chondrules and matrix.

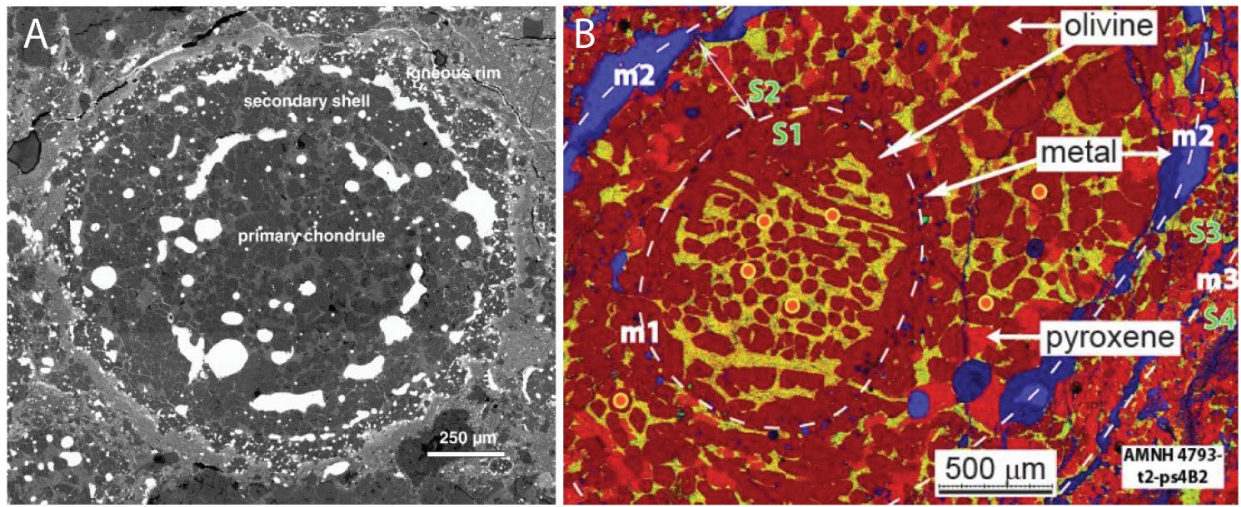

*Fig. 3: Chondrules exhibiting complex structures with multiple zoning and rims, which is indicative of multiple melting events. **A**) Chondrule from the LAP 02342 CR2 chondrite with a primary and secondary shell and a surrounding igneous rim, all separated by metal rims (from Rubin, 2010). **B**) Complex chondrule from CR2 chondrite Acfer 139 with four shells (s) and metal (m) rims (from Hobbart et al., 2015), with Si = red, Ca = green,*

*Fe = blue. Olivine in pyroxene-bearing layer S2 grew epitaxialy on olivine in pyroxene-free layer S1, the two separated by a thin layer M1 of numerous metal grains.*

### 2.5.2 FeO-poor, FeO-rich and Al-rich chondrules

Chondrules contain many different textures, petrologies and compositions and it is not well understood whether these variations are attributable to local or large-scale differences in the chondrule-forming environments. The observed complementarity includes a non-solar bulk matrix composition that is offset by a bulk chondrule composition, which includes FeO-poor type I chondrules, the less abundant FeO-rich type II and Al-rich (>10 wt.% Al) chondrules (Ebel et al., 2008). Within the complementarity model, this requires all chondrules to form together from the same local reservoir. Alternatively, it is, for example, possible that the subsolar matrix is complementary to a supersolar chondrule population dominated by FeO-poor chondrules (Hezel et al., 2010). If these are mixed with hypothetical solar FeO-rich chondrules, this will produce 'hybrid complementarity' (Jacquet et al., 2016; Goldberg et al., 2015). If correct, a significant portion of a chondrite (<10 % of carbonaceous and <50 % of ordinary) is formed independently of the complementary chondrule and matrix components. It has not been well established whether FeO-poor and FeO-rich chondrules formed co-genetically. They have been proposed to have formed together in heterogeneous formation regions with sizes that limited diffusive equilibration on chondrule formation timescales (i.e, redox conditions; Cuzzi and Alexander, 2006; Libourel et al., 2023) or sequentially (i.e., relict Mg-rich olivine found in FeO-rich chondrules, Connolly and Huss, 2010; Villeneuve et al., 2015; Wasson and Rubin, 2003; Jacquet et al., 2015). For Al-rich chondrules, it is generally accepted that the compositions of these chondrules reflect the addition of refractory inclusions to the precursor inventory (Krot and Keil, 2002). The inference of similar thermodynamic formation conditions between ferromagnesian and Al-rich chondrules has been used to conclude that these objects formed during the same or at least similar heating events as typical chondrules (Tronche et al., 2007; Zhang et al., 2020). If Al-rich chondrules formed co-genetically with ferromagnesian chondrules, this requires a nugget effect in the precursor dust from which the chondrules formed (i.e., a random distribution of Al- and Ca-rich refractory grains). In contrast, co-genetic relationships between FeO-rich and FeO-poor chondrules could be established by O and nucleosynthetic isotope studies of these objects since their compositional variations are thought to result from environmental conditions rather than nugget effects in the precursor inventory. In sum, the evidence presented so far suggests that the petrological variation of chondrule populations poses no restriction in the complementarity model.

## 3. Evidence from the chemical compositions of chondrules and matrix

To apply Baldy's Law, the idea that "some of it plus the rest of it equals all of it", to the chemistry of chondrites and their components requires knowledge of the bulk compositions of the chondrite groups, the relative fractions of "clasts" (mostly chondrules) and matrix among the different groups, and knowledge of the average compositions of the clasts and matrix in each group. There is a large literature on bulk chondrite compositions that is reviewed elsewhere (Lodders, 2021). Recent work has refined the measurement of component abundances as fractions of total area of polished sections. Analysis of individual chondrule and bulk matrix is challenging and the source of debate. The compositions of bulk chondrules and matrix are different for most elements studied so far.

Differences in element compositions are often compared as element ratios, as these are independent from absolute element concentrations that can be modified by variations in the abundances of major elements.

## 3.1 Bulk major element compositions of chondrites

Chondrites have bulk compositions that for some element ratios are similar to CI chondritic compositions. This is primarily the case for the more refractory elements, such as Mg/Si, Ca/Al, or Ti/Al. Other majorn elements, such as Fe, are readily disturbed by terrestrial weathering. A number of carbonaceous chondrite groups have nearly CI-like Mg/Si ratios, whereas their bulk chondrule and matrix compositions do not have the CI-like Mg/Si ratios. This constellation – chondrules and matrix with different element ratios, while the bulk chondrite has CI-like ratios – seems an unlikely outcome of mixing two reservoirs – i.e., a chondrule and a matrix reservoir – with different compositions (Fig. 1). This special case of different chondrule-matrix compositions with at the same time about bulk chondrite CI chondritic compositions has been termed 'complementarity' throughout the literature. This complementarity has so far been reported primarily for carbonaceous and R chondrites. In this section, we focus on the Mg/Si ratios of bulk chondrites, chondrules and matrix, since other elements such as Ca, Ti and Al can be affected by contributions from CAIs or by redistribution during secondary alteration (i.e., Fe, S and Ca; see section 5).

## 3.2 Bulk chondrite compositions from volatility patterns

While the bulk chemical compositions of chondrites, by group, are quite similar, they differ from the Solar bulk composition as approximated by CI chondrites (Lodders and Fegley, 1998; Braukmüller et al., 2018; Lodders 2020, 2021). Grossman (1996) argued that each chondrite group accreted from a distinct reservoir established prior to chondrule formation and prior to chondrite accretion, a statement of the single-reservoir hypothesis. As noted by multiple authors (most recently by Braukmüller et al. 2018), bulk carbonaceous chondrites show roughly monotonic decreases in elemental abundances, relative to the CI composition, with increasing volatility (i.e., their 50% condensation temperatures under canonical conditions, Lodders, 2003) until at condensation temperatures of ≤800 K their abundances plateau. These depletion patterns had previously been interpreted as the product of a partial condensation process in the inner Solar System with a radial thermal gradient, where chondrites formed in distinct reservoirs reflected by the degree of their (moderately) volatile element depletions (Wasson and Chou, 1974; Wai and Wasson, 1977). Some of these models have postulated these depletion patterns to be the result of mixing of chondrules with volatile depleted compositions and matrices with CI-like compositions (e.g., Anders, 1977; Alexander, 2005,2019). Seemingly ruling out this model, the bulk matrix compositions of many carbonaceous chondrites were found to be volatile-depleted, relative to CI chondrites (Bland et al., 2005). However, matrix analyses of relatively unaltered carbonaceous chondrites are much more CI-like (Hewins et al., 2014; Zanda et al., 2018; van Kooten et al., 2019), unlike their more altered counterparts (Bland et al., 2005; van Kooten et al., 2019). This suggests that the matrices of carbonaceous chondrites at the time of accretion were roughly CI-like, which would be consistent with presolar grain and organic abundances, and that they were not subject to heating and volatile loss during chondrule formation. Alteration may not be the only cause of deviations from CI. The addition of chondrule and refractory inclusion materials to the primordial matrix, as defined here, would also cause apparent fractionations in analyzed areas (Alexander et al., 2005; Bland et al. 2007; Jacquet et al., 2016; Braukmüller et al., 2018). If the matrices of chondrites were CI-like at the time of accretion, it requires that the matrix come from some distant (not quantified) reservoir. Nevertheless, even if the bulk matrix did not experience (moderately) volatile element loss relative to CI chondrites, as we will discuss below this component undoubtedly has a sub-chondritic Mg/Si ratio that has been connected to chondrule-matrix complementarity.

### 3.3 Chondrule/matrix component abundances

To assess whether the chemical compositions of chondrite components are complementary in each group, it is important to know the correct abundance ratios of chondrules, refractory inclusions, and matrix in each of the chondrite classes. There is relatively little original data on this topic, most based on optical point counting. In Figure 4, we summarize the recently published results for CM (Fendrich and Ebel, 2021; Patzer et al. 2023), CO and CV (Ebel et al., 2016; Patzer et al. 2021), CR (Ebel et al., 2024; Patzer et al. 2022), K (Barosch et al., 2020), and LL chondrites (Ebel et al., 2024)These data are based on either image analysis of stacked (registered) X-ray maps of large areas of thin and thick sections, or point-counting of thin sections with an electron microprobe. Results for the abundances of chondrules, matrix and refractory inclusions agree, within the uncertainties and variability within meteorites and groups, with earlier studies (e.g., reviews by Weisberg et al., 2006, based on Grossman, 1988, citing McSween, 1977ab, 1979 and Huss et al., 1981). The broad literature remains in agreement on the central point, that the different chondrite classes have highly variable chondrule/matrix ratios.

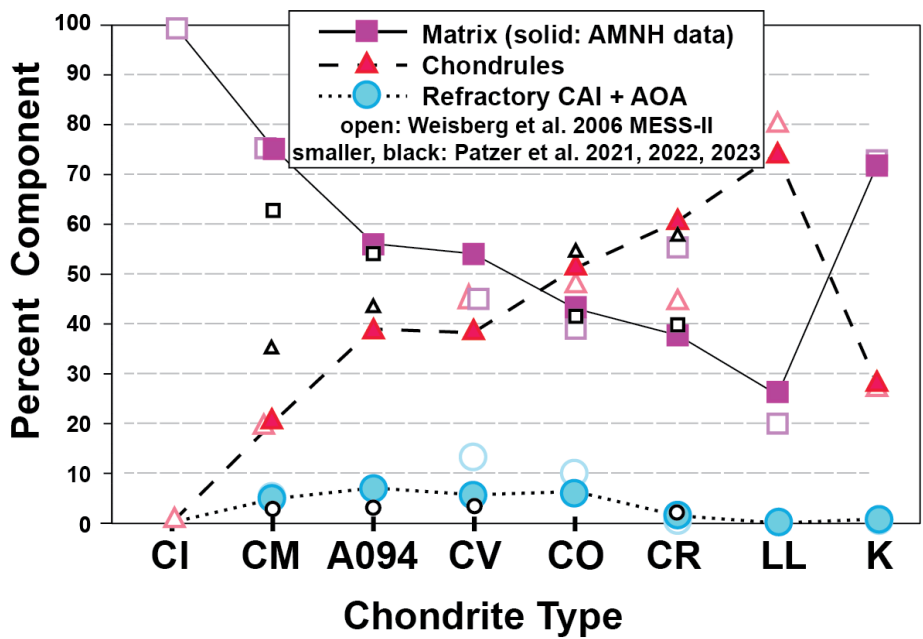

*Figure 4: **The percent component of chondrules, refractory inclusions and matrix by surface area** in CI, CM, Acfer 094, CV, CO, CR, LL, and K chondrites (references in text). Lines start at CI (99% matrix). Data is compared to Table 1 of Weisberg et al. (2006, large open symbols) and Patzer et al. (2021, 2022, 2023; smaller, black open symbols, with Renazzo data only, for CR). Note that the percent component of chondrules and matrix in CM*

*chondrites from Patzer et al. (2023) are for the least altered CMs, removing much of the uncertainty associated with estimating matrix abundance the CMs. Adapted from Ebel et al. (2024, their Fig. 9).*

### 3.4 The Mg/Si ratios

In this section, we will explore the Mg and Si elemental abundances of chondrules and matrix relative to the bulk chondrite and CI chondrites. For individual data points and methods used, the reader is referred to the supplementary data Table **S1**.

*3.4.1 The matrix*
Hezel and Palme (2010) and references therein observed subsolar Mg/Si ratios in the matrix of CV, CR, CO and CM chondrites. They noted that "absolute concentrations from broad beam matrix analyses are unreliable as matrix can be porous and contains sulfide and metal" which are calculated as FeO, but that "the Mg/Si ratios are unaffected by these analytical limitations". Similarly, later research on chondrule and matrix compositions from relatively unaltered CO, CM, CV and CR chondrites have found chondritic to subsolar Mg/Si ratios of the matrix (Hewins et al., 2014; Ebel et al., 2016, 2024; van Kooten et al., 2019; Patzer et al., 2021, 2022, 2023). The averaged matrix compositions of carbonaceous chondrite groups are very similar and all appear to be richer in Si relative to CI chondrites (**Fig. 5B**). It has been questioned whether this increase reflects a systematic analytical error or if it is a real feature of the matrix. Zanda et al. (2018) compared results of different analytical methods for measuring compositions of matrix and chondrules. They pointed out that "caution needs to be exercised when data obtained from different analytical methods are used as a basis for understanding chondritic compositions", based on a comparison of broad beam EPMA on Orgueil with earlier wet chemical results. One reason for caution is that the most commonly used technique, EPMA, assumes that the sample is physically and chemically homogeneous, which is not the case for the porous and polyphase matrix. Other techniques may also suffer from artifacts associated with the analysis of the polyphase matrix. Zanda et al. (2018) found that their Mg/Si wt.% ratio for Orgueil measured by EPMA was 10-20% lower than that measured by bulk chemical methods. Comparison with CI chondrites is an extreme case, as these are the most hydrous, volatile-rich, and most porous of all chondrites (Macke et al., 2011). Huss et al. (1979, 1981) attempted to test the accuracy of broad-beam EPMA of matrix by analyzing a pressed pellet of powdered peridotite as a proxy for ordinary chondrite matrix. The normalized analysis of the pellet agreed with the published data, except for the MgO (7% too high) and the S concentration (25% too low). Another comparison was made between broad-beam EPMA (Kimura and Ikeda, 1998; McSween and Richardson, 1977) and wet chemical analysis (Clarke et al., 1970) for Allende matrix (**Table 1**). They noted that "weight per cent Mg/Si in matrices is constant (0.82 ± 0.05) but less than ratios derived from bulk analyses". McSween and Richardson (1977) compared the EPMA Allende matrix analyses to wet chemical analysis of a <100 μm sieved fraction by Clarke et al. (1970), reporting that "agreement between the two methods of analysis is generally very good". However, a <100 μm fraction does not meet the definition of matrix of <5 μm given by Scott et al. (1988). Moreover, Allende is a hydrothermally altered CV chondrite that has probably lost porosity relative to unaltered chondrite matrix (Krot et al., 1995). Nevertheless, at face value, the matrix Mg/Si ratio obtained by broad-beam EPMA is higher than that obtained by wet analysis (Table 1), where we would expect an opposite result if we were to explain the subsolar matrix as an analytical artifact. Hence, we conclude that carbonaceous chondrite matrix is in fact subsolar for its Mg/Si ratio. Although the data at hand does not favor an analytical artifact as a cause for subsolar Mg/Si ratios, a comprehensive comparison between broad-beam and focused spot EPMA, LA-ICP and wet chemical analyses of the various chondritic matrices (altered and unaltered) is needed to understand the potential caveats for using these techniques.

*Table 1: Comparison of Si, Mg and Al abundances in bulk and matrix of Allende (CV3) for three techniques. Data are presented as oxides and recalculated to atom. wt.%. Broad beam EPMA of Allende matrix yielded a higher Mg/Si ratio than the only wet chemical analysis available (Clarke et al., 1970, analysis by Jarosewich) but a lower Al/Si ratio. Data are from Clarke et al. (1970, C+ 70), Jarosewich et al. (1987, J+ 87), Stracke et al. (2012, S+ 12), McSween and Richardson (1977, MR 77), and Kimura and Ikeda (1998, KI 98).*

|  | wet chemistry | | | XRF | broad beam EMPA | |
|---|---|---|---|---|---|---|
|  | C+70 bulk | J+87 bulk | C+70 matrix | S+12 bulk | MR77 matrix | KI98 matrix |
| SiO$_2$ | 34.23 | 34.27 | 33.11 | 34.24 | 28.00 | 27.90 |
| MgO | 24.62 | 24.51 | 21.42 | 24.77 | 20.20 | 19.00 |
| Al$_2$O$_3$ | 3.27 | 3.35 | 3.07 | 3.00 | 2.30 | 2.10 |
| Si | 16.00 | 16.02 | 15.48 | 16.00 | 13.09 | 13.04 |
| Mg | 14.85 | 14.78 | 12.92 | 14.94 | 12.18 | 11.46 |
| Al | 1.73 | 1.77 | 1.62 | 1.59 | 1.22 | 1.11 |
| Mg/Si | 0.928 | 0.923 | **0.835** | 0.934 | **0.930** | 0.879 |
| Al/Si | 0.108 | 0.110 | 0.105 | 0.099 | 0.093 | 0.085 |

### 3.4.2 Chondrules

The diversity of compositions among chondrules in the carbonaceous chondrites is much larger than seen in ordinary chondrite chondrules (Lobo et al., 2014, **Fig. 5A**). Chondrules generally contain higher proportions of Mg and Si than matrix in all groups, but their diversity makes it difficult to obtain the statistical power to constrain their aggregate or mean composition in any chondrite.

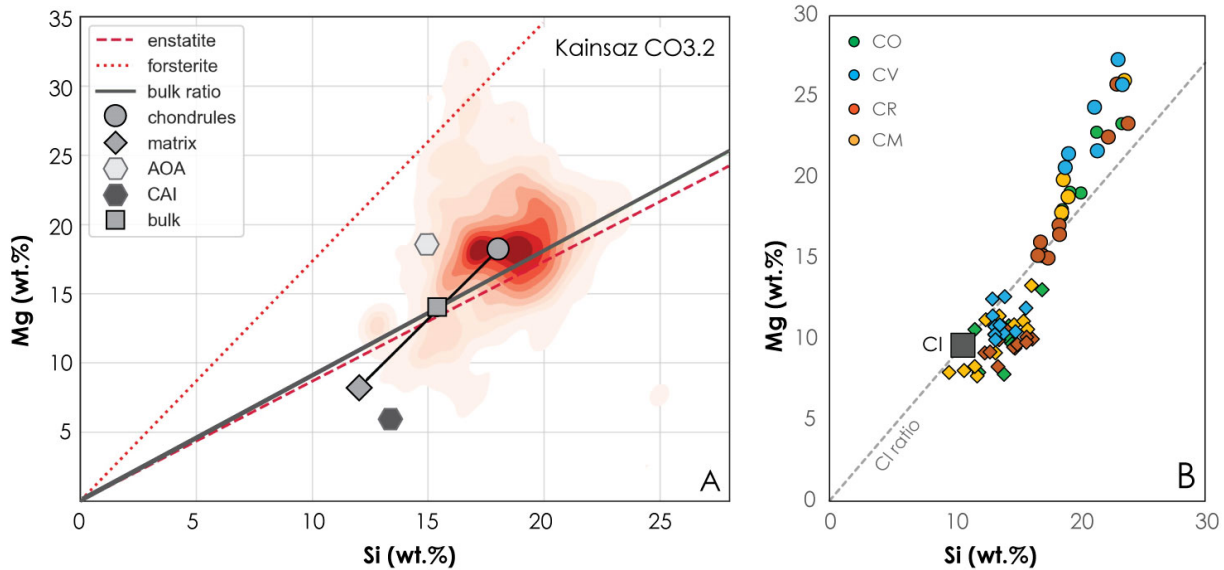

*Fig. 5: Si and Mg contents of chondrules, matrix and bulk for CO, CM, CR and CV chondrites. A) Weighted density distribution plot for CO3.2 chondrite Kainsaz, using 3367 chondrules from Ebel et al. (2016). Isolated olivine grains in the matrix are counted as chondrules and metal grains in the matrix are counted as matrix. B) Compiled Si and Mg contents for averaged chondrules and matrix from individual CO, CV, CR and CM chondrites with data from Hezel and Palme (2010 and references therein), Hewins et al. (2014) and Patzer et al. (2021, 2022, 2023). See supplementary table 1 for data and methods used. Diamonds are matrix, spheres are chondrules and squares are bulk analyses.*

A wide variety of EPMA techniques have been used to analyze chondrule compositions (see supplementary data **table S1**). Hezel and Palme (2010) calculated bulk chondrule compositions (10 in CO, 13 in CR, 8 in CV and CM) by modal reconstruction from broad beam EPMA and modes derived from element maps. An "absolute error" correction was applied to 2D data based on Hezel and Kießwetter (2010). Such an approach may not be representative of the Mg/Si ratio in chondrules, since surficial sections may be more pyroxene-rich with Mg/Si ratios closer to unity, than equatorial sections. Ebel et al. (2016) analyzed many more chondrules using spot EPMA analyses to measure bulk compositions across thousands of chondrules in CO and CV chondrites, and millions of matrix spots, corrected in **Figure 5A** for Kainsaz (CO3.2) using calibration curves on standards (Crapster-Pregont et al., 2020). For CO chondrites, the Mg/Si ratios of the matrix are consistently more CI-like with increasing petrological grade, whereas the chondrules show the opposite effect (Ebel et al., 2016, their Fig. 10). Similar to the Mg/Si ratios of the matrix, the overall supersolar Mg/Si ratios of chondrules in carbonaceous chondrites are observed by various authors and through different methods (see supplementary data **Table S1**). It is unclear to what extent the density differences between different chondrule phases (i.e., olivine, pyroxene, mesostasis, sulfide and metal) influence the Mg and Si concentrations determined by EPMA broad-beam analyses. For example, Hezel and Palme (2010) commented that an increasing amount of metal/sulfide in the chondrule results in higher concentrations of Mg and Si. This may be reflected in the relatively higher Mg and Si concentrations of CV chondrules (**Fig. 5B**), which typically contain more metal/sulfide within the chondrules. However, although metal and sulfide in the chondrules affect the absolute Mg and Si concentrations, the Mg/Si ratio should remain constant. Therefore, the offset in Mg and Si from the CI-line as observed in **Figure 5**, is not an analytical artifact. As commented above, the huge spread of chondrule compositions makes it that much harder to average and compare the composition of the chondrules with that of the matrix and the bulk with the aim to make a significant statement about chondrule-matrix complementarity. Where statistical power is possible, the "gold standard" of this kind of work is the reproducibility of chondrite bulk compositions from determination of the abundances (area fraction), and compositions of many refractory inclusions, chondrules, and matrix. In the next section, we will discuss these results and whether they point to a genetic relationship between chondrules and matrix.

### 3.4.3 Interpreting the Mg/Si ratio of chondrites and their components
As discussed above, the Mg/Si ratios of CV, CR, CM and CO chondrite matrices appear to be subsolar and this is a primary feature. However, the overall chemical compositions of these matrices are CI-like and do not exhibit significant volatile loss or gain (see section 3.3). At the same time, in contrast to the matrices, there is a large spread in the Mg/Si ratios of chondrules. Chondrule formation is generally accepted to be an open system process, where interaction with the gas determines a large fraction of the chondrule chemistry (Friend et al., 2016; Barosch et al., 2019). For example, the mineralogical zonation in chondrules can be readily explained by the progressive condensation of SiO gas into the chondrule melt since SiO remains longer in the gas phase compared to other major elements (Ebel et al., 2000). As such, the chondrule composition reflects the shift from a nominally olivine to enstatite composition and a corresponding decrease in the Mg/Si ratio of a chondrule (Friend et al., 2016). Hence, the range of Mg/Si ratio in chondrules is likely to reflect this melt-gas interaction and appears to be a generic feature of chondrule formation. This results in bulk chondrule populations from different chondrite groups (including CR, CO, CM and CV) with supersolar to solar Mg/Si ratios between the forsterite and enstatite correlations (**Fig. 5A**). If the SiO gas condenses onto the matrix that is more coupled to the gas, the initially CI-like matrix becomes subsolar for Mg/Si.

Another interpretation that has been put forward to explain a subsolar Mg/Si in matrix is the loss of a component that stoichiometrically resembles forsterite (Hezel et al., 2018; Patzer et al. 2022). If this forsteritic component were then incorporated into chondrules or their precursors prior to or during their formation, this could explain supersolar Mg/Si ratios of CV chondrules (Hezel et al. 2018). It is unclear what kind of physical process can be invoked to produce this complementarity (see section 6.5 and 7), especially since the depletion in the matrix is restricted to forsterite and does not involve loss of volatiles relative to the CI starting composition. Patzer et al. (2022) proposed that the matrix could have lost a forsterite component by aerodynamic sorting, given that large cluster IDPs contain large forsterite grains (adding up to roughly solar compositions) that are absent in smaller IDPs with supersolar volatile compositions. Alternatively, apparent forsterite-enrichments in chondrules (and depletions in matrix) have been interpreted as cases of 'mistaken identity', where forsterite grains in the matrix have been counted as chondrule fragments, even though they are related to the matrix fraction or represent small AOA-like fragments (Russell et al., 2010). Distinguishing chondrule-related forsterite (i.e., type I chondrule phenocrysts) from olivine grains in the matrix that were not part of a chondrule is no straightforward process, since both types can have compositional relationships to AOAs (Jacquet and Marrocchi, 2017). Recent X-ray imaging of forsterite grains in the Murchison matrix has shown co-genetic relationships of these grains with chondrules (Perrotti et al., 2021), in agreement with O isotope and trace element analyses of these fragments (Jacquet et al., 2021 and references therein). If we can extrapolate these results to most ~5 μm forsterite grains, the 'mistaken identity' hypothesis does not hold, since removing these forsterite grains from the matrix counts will only result in a more subsolar Mg/Si ratio of the matrix. This implies that a better understanding of the origin of micron- to submicron-sized forsterite (and Mg-rich pyroxene) grains located in the matrix can lead to tighter constraints on genetic relationships between chondrules and matrix.

Alternative solutions to subsolar Mg/Si ratios for the matrix do not require a complementary relationship with chondrules. CI-like dark clasts from polymict ureilites, HED meteorites, ordinary and CR chondrites also contain solar to subsolar Mg/Si ratios (**Fig. 6**; Patzek et al., 2018). These clasts do not contain chondrules and some of these clasts have proposed outer disk origins (Nittler et al., 2019; Kebukawa et al., 2019, van Kooten et al., 2024), showing a continuum in chemically CI-like dust with different (nucleosynthetic) isotope compositions (van Kooten et al., 2024). As such, the changes from solar to subsolar Mg/Si ratios could reflect a change in the dust composition of the outer disk (**Fig. 6**).

***Fig. 6:*** *The MgO versus SiO$_2$ contents of CI-like dark clasts from CR chondrites, polymict ureilites, HED meteorites and ordinary chondrites (data from Patzek et al., 2018). These clasts do not contain chondrules and show solar to subsolar Mg/Si ratios. Bulk GEMS compositions (Keller and Messenger, 2011; Ishii et al., 2008) show the same subsolar Mg/Si ratios (plotted using a CI chondrite MgO content) as CI-like clasts from CR chondrites, indicating a change in outer disk material (van Kooten et al., 2024).*

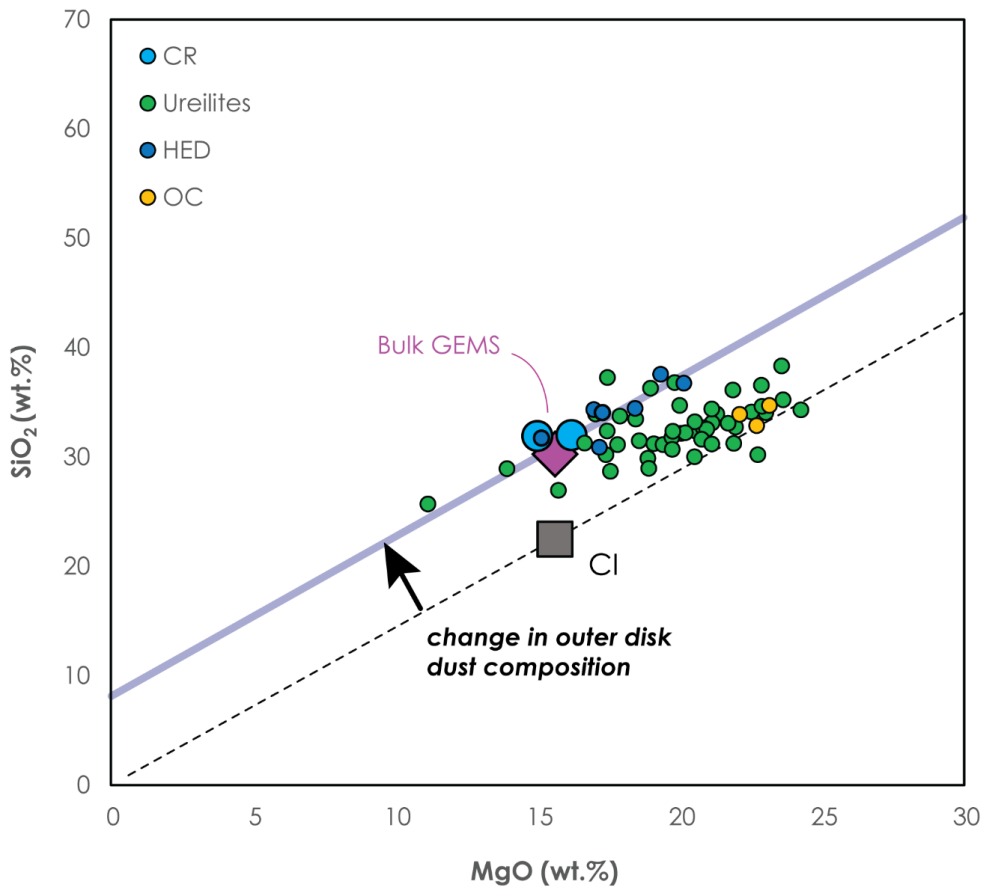

In summary, a few key areas emerge on which we should focus our attention to correctly interpret the Mg/Si ratios of chondrules and matrix. First, although it appears that preliminary tests comparing EPMA with wet chemical analyses of matrix cannot explain its subsolar Mg/Si ratio, further research is needed to systematically understand the effects of porosity and polyphase mineralogy that can offset EPMA from wet analyses data. Secondly, the broad definition of matrix used throughout literature (grains <5–100 μm) can include or exclude forsterite (but also pyroxene, metal, and sulfide) grains that have a poorly defined origin. A deeper understanding of the relation of these grains to chondrules or matrix is critical to define the final Mg/Si ratios of these components. Finally, a detailed micron- to nanoscale investigation of these grains in pristine chondrites and their potential relationship to other primary matrix components can help constrain their abundance and origin and their importance in establishing subsolar Mg/Si ratios in the matrix of carbonaceous chondrites.

## 4. Evidence from mass-dependent and mass-independent isotope systematics

Mass-dependent isotope variations can help us understand the physical processes of evaporation and condensation in the chondrule forming reservoir, with a side note that these processes with respect to chondrule formation are not yet fully understood (Hewins et al., 2005; Jones et al., 2018). For example, for free evaporation of moderately volatile elements (MVEs) from chondrules during high temperature melting events, an isotopically heavy chondrule residue is predicted (Ebel et al., 2018). However, for multiple isotope systems (i.e., Zn, K, Rb, Te, Ga, Cu) we observe (or infer from bulk compositions) isotopically light chondrules, relative to CI (Hellmann et al., 2020; Nie et al., 2021; Pringle et al., 2017; Pringle and Moynier, 2017; van Kooten and Moynier, 2019, Kato and Moynier, 2017; Luck et al., 2003, 2005; Hu et al., 2023), which is usually interpreted as either physical loss of chondrule melt components (i.e., sulfide), interaction with an isotopically light gas (i.e., SiO) or kinetic isotope fractionation during condensation from an isotopically 'normal' gas (Alexander et al., 2022). Moreover, the presence of Na, K, S, and Fe in chondrules in some chondrites all point to their formation in relatively dense environments that enabled chondrule melts to equilibrate with their vapor after relatively little evaporation (Alexander et al., 2022, 2008).

In addition to mass-dependent variations, mass independent isotope variations are also observed in chondritic components. These isotope signatures represent nucleosynthetic variations that are independent of any naturally occurring mass-dependent isotope fractionation processes (i.e., evaporation, condensation, fractional crystallization, etc.). As such, variations in mass independent isotope signatures are generally thought to reflect spatio-temporal changes in the dust isotope composition of the protoplanetary disk that are the result of incomplete mixing or unmixing of dust. For example, the distinct $\mu^{54}$Cr values of bulk non-carbonaceous and carbonaceous chondrites are generally accepted to represent accretion of their parent bodies at different times and orbital distances from the protoSun (Trinquier et al., 2009, 2007; Warren, 2011). Ungrouped achondrites have been linked to chondritic counterparts by their similar Cr isotope signatures (Sanborn et al., 2019). Although having the same nucleosynthetic signatures in all elements does not require an origin from the same parent body, it does imply formation within the same isotopic reservoir (even though there are no clear spatial constraints on these reservoirs).

Extrapolating to chondritic components, we can infer a genetic link between chondrules and matrix if these components possess *identical* nucleosynthetic isotope signatures, assuming the matrix and chondrules formed from the same precursor material. This is because if chondrules formed from the same dust inventory as the matrix, processes such as evaporation and/or condensation will not erase the mass independent signature. The only way for chondrules to have changed their nucleosynthetic compositions prior to or during chondrule formation would have been by physically gaining or losing (without equilibrating with the remaining melt) a mineral phase(s) that was a carrier of a nucleosynthetic anomaly.

*Fig. 7: **Mass-dependent isotope compositions** of chondrules (spheres), matrix (diamond) and bulk chondrites (square) for CR (blue), CV (yellow) and CM (gray) chondrites. Data is compiled for Fe (Hezel et al., 2010, 2018b), Mg (Olsen et al., 2016), Si (Kadlag et al., 2021), Zn (van Kooten and Moynier, 2019) and Te (Hellmann et al., 2020). We also show the non-matrix fraction (i.e., chondrules and refractory inclusions) calculated by Hellmann et al. (2020). CI chondrite isotope compositions are $\delta^{30}Si_{CI}$ = −0.48±0.01 ‰ (Onyett et al., 2023), $\delta^{25}Mg_{CI}$ = 0.01±0.04 (Larsen et al., 2011; Bizzarro et al., 2011), $\delta^{56}Fe_{CI}$ = 0.06±0.03 ‰ (Schiller et al., 2020), and plot at the same locations as CM, CV and CR bulk chondrites. $\delta^{128}Te_{CI}$ = 0.15±0.04 ‰ (Hellman et al., 2020) and $\delta^{66}Zn_{CI}$ = 0.48±0.06 ‰ (Luck et al., 2005; Pringle et al., 2017) are typically enriched relative to bulk chondrites.*

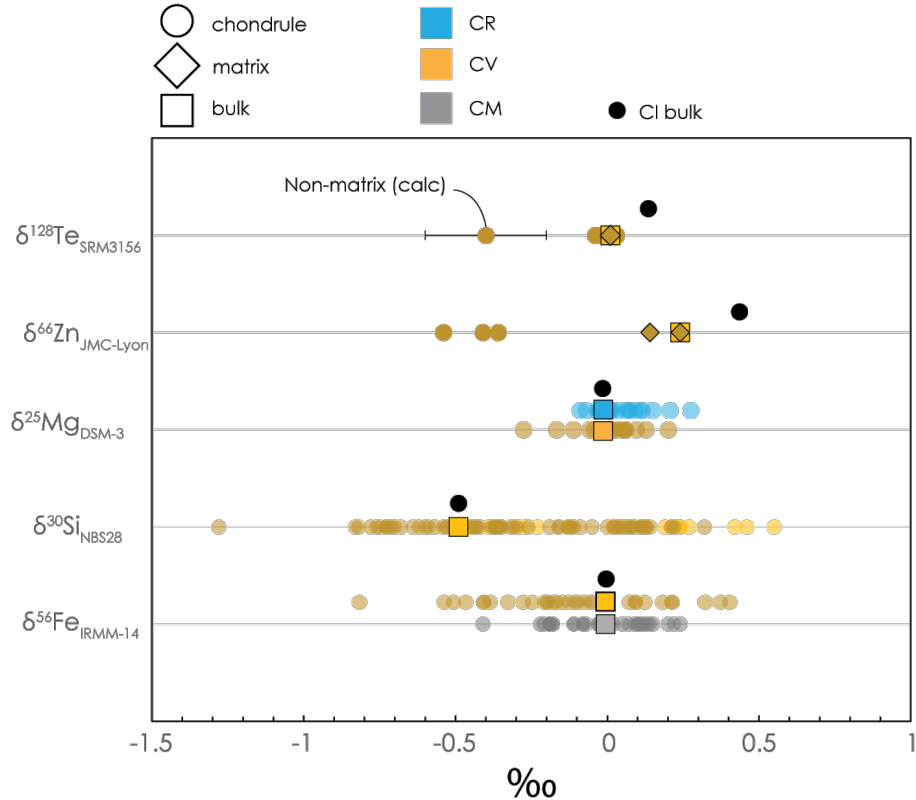

## 4.1 Mass-dependent isotope signatures

Provided that there was relatively unrestricted evaporation (e.g., Rayleigh-fractionation-like behavior), mass-dependent heavy isotope enrichments of chondrules are predicted (Ebel et al., 2018). This is particularly true for (moderately) volatile elements (i.e., those with 50% condensation temperatures below ~1350 K), based on their elemental depletions, relative to CI, in chondrules from CV, CO, CM and CR chondrites (Kong and Palme, 1999; Rubin and Wasson, 1988, 1987; van Kooten et al., 2019). Significant elemental depletions, relative to CI, are also observed for Fe and other siderophile elements such as Ni and Co, which largely reflects physical metal/sulfide loss from the chondrules (which may have been partially retained as metal rims around chondrules). Major elements Mg, Fe and Si, which have relatively high condensation temperatures (~1300 K, Ebel, 2006), have been analyzed for their mass-dependent isotope compositions in bulk aggregates of separated chondrules and individual chondrules from CV, CM and CR chondrites. For these elements, there are modest ranges of mass-dependent isotopic variations that are both positive and negative relative to the terrestrial value and CI chondrites (**Fig. 7**; Bouvier et al., 2013; Hezel et al., 2010, 2018b; Kadlag et al., 2021; Olsen et al., 2016; van Kooten et al., 2020). For some Al-rich chondrules, there is a clear contribution of refractory inclusions shifting the $\delta^{25}$Mg of the chondrules to more extreme values (Bouvier et al., 2013; Olsen et al., 2016). For Mg isotopes, no correlation is observed between the elemental compositions and textures of the chondrules (barred versus porphyritic), and the Mg isotope variability (Olsen et al., 2016). The small variations in $\delta^{25}$Mg values, but lack of correlation with Al/Mg ratios, in CV and CR chondrules have been suggested to either reflect 1) complex histories of evaporation and condensation for the chondrules (Olsen et al., 2016) and/or subsequent secondary alteration (Bouvier et al., 2013), or 2) the accretion of precursor materials with variable $\delta^{25}$Mg values (Olsen et al., 2016). The former scenario (i.e., evaporation/condensation) is also favored to explain the Fe isotope data of bulk CV and CM chondrules (Hezel et al., 2010, 2018b). Here, the added complexity is that besides possibly experiencing some evaporative Fe loss, chondrules also experienced metal/sulfide loss, which can potentially create opposing directions in isotopic fractionation. Silicon isotope variations in CV chondrules have also been interpreted as the result of an evaporation process, complicated by the interaction of chondrules with an isotopically light gas (Kadlag et al., 2021). Collectively, these results point towards a generic chondrule formation process, where the bulk reservoir of all chondrites or the individual reservoirs of each chondrite group, had a similar isotope composition of these elements to CI chondrites, which was then subjected to the chondrule forming process described above, creating negative and positive isotope fractionations relative to the bulk value.

Given the depletion of MVE abundances in CC chondrules (and bulk chondrites), relative to CI, mass-dependent isotope fractionation of MVEs such as S, Zn, Cu, Rb, K and Te are expected to be large if they were lost via free evaporation (up to 10s of permil, depending on the degree of depletion). The potential levels of mass fractionation are large enough that, if present, they should be seen in the isotopic compositions of bulk chondrites. This is the case even if the more volatile-rich matrix dilutes the signal, unless the matrix is the complement to the chondrules. In fact, no systematic heavy isotope enrichments that correlate with MVE elemental depletions have been reported in bulk chondrites. On the contrary, for many, but not all MVEs, their bulk isotopic compositions are light isotope enriched, relative to CI. The lack of such large heavy isotope deviations in the bulk chondrite compositions, as well as in some cases the observed or inferred light MVE isotope signatures of chondrules, have been proposed to reflect 1) equilibrium between chondrules and surrounding gas near peak formation temperatures to suppress heavy isotope fractionation, and 2) kinetic light isotope fractionation due to incomplete recondensation during cooling, and/or loss of MVE-bearing metals and sulfides (Pringle et al., 2017; van Kooten and Moynier, 2019).

***Fig. 8:*** *The mass-dependent isotope ratios in delta notation of Zn (Luck et al., 2005; Pringle et al., 2017), Rb (Nie et al., 2021) and Te (Hellmann et al., 2020) from chondrites with different matrix fractions in weight percent (from Alexander, 2019). Errors represent 2SD of the mean value. We also plot the average CV chondrule composition (van Kooten and Moynier, 2019). Note that a mixing line between the matrix and chondrule endmember is hyperbolic, since chondrules have a much lower Zn concentration (<5 ppm) than the matrix (~300 ppm).*

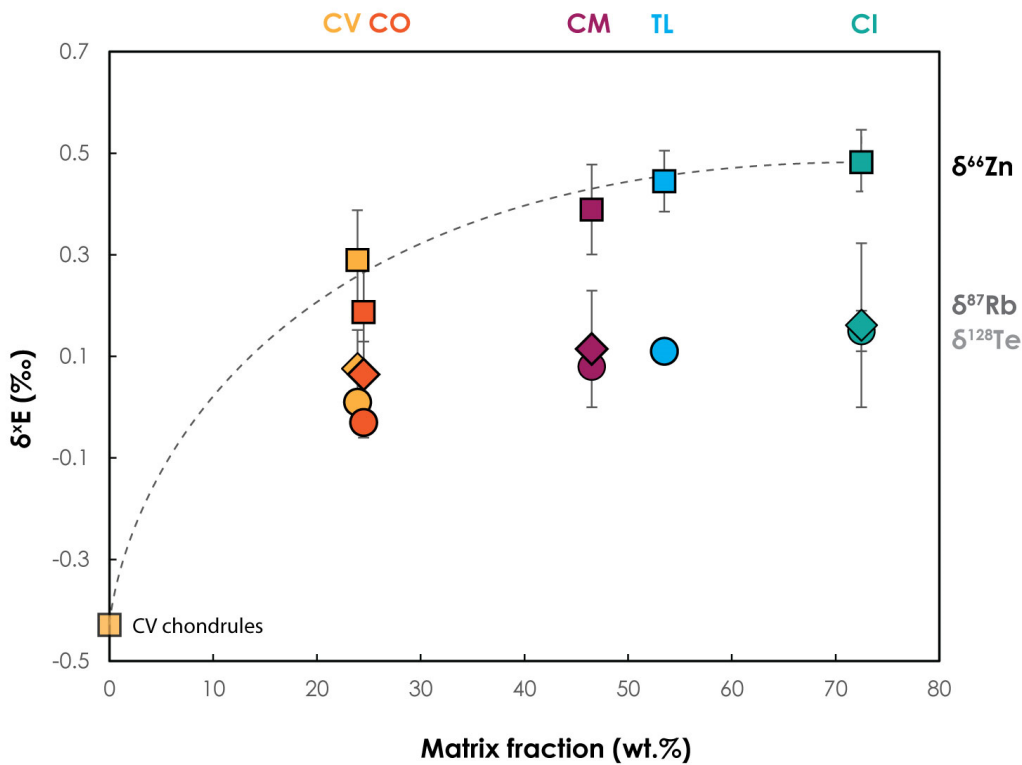

Many of the MVEs are potentially mobile during hydrothermal alteration and thermal metamorphism. Thus, measurements in situ or of physically separated components must be interpreted with caution (see section 5). For instance, the K isotope systematics of chondrules, matrix and CAIs in Allende (CV3.6) can be mainly attributed to secondary processes (Jiang et al., 2021). No systematic heavy K isotope enrichments that correlated with K depletions were found in Semarkona (LL3.00) chondrules after selection of samples with the least petrologic evidence for K exchange between chondrules and matrix (Alexander and Grossman, 2005). However, a separate study of Semarkona chondrules found that all the measured chondrule glasses had exchanged O isotopes with the matrix (Kita et al. 2010). That chondrules in a type 3.00 have extensively exchanged O isotopes with the matrix should give one pause when studying samples of higher petrologic type. Sulfur appears to have been relatively immobile at least in the matrix of Semarkona (Grossman and Brearley, 2005). Tachibana and Huss (2005) measured S isotopes in Semarkona chondrules and found no evidence for significant isotopic mass fractionation, although alteration of metal/sulfide grains in many chondrules suggests that they may have been open to S exchange with the matrix too. Similar observations were made for Zn isotopes in chondrules from Allende (CV3.6) and the less altered but shock heated Leoville (CV3.1). Zinc isotope systematics from Allende are more variable and show higher $\delta^{66}$Zn values that are closer to the bulk CV (Pringle et al., 2017) than in Leoville, which contains chondrules with consistently light Zn isotope signatures (van Kooten et al., 2021; van Kooten and Moynier, 2019). The Zn isotope compositions from Leoville chondrules are interpreted as primary signatures (or at least as upper limits) and agree with Te isotope signatures calculated for the non-matrix fraction of CV chondrites (**Fig. 7**, Hellmann et al., 2020). The Zn isotope composition of the Leoville matrix fraction is indistinguishable from the bulk value (van Kooten and Moynier, 2019).

A different approach for inferring chondrule and matrix compositions that is less sensitive to internal parent body redistribution of elements is to correlate bulk chondrite mass-dependent isotope compositions with estimates of their chondrule and matrix fractions. For Zn, Cu, K, Rb, Te, Sn and Ga, the bulk carbonaceous chondrite isotopic compositions are best explained by two-component mixing of a chondrule endmember consisting of isotopically light Zn, Cu, K, Rb, Te, Sn and Ga, and a CI-like matrix endmember (Luck et al. 2003, 2005; Nie et al., 2021, 2023; Hellmann et al. 2020, 2023; Hu et al. 2023; Koefoed et al. 2023) (**Fig. 8**). This is consistent with combined chemical, Zn and Cr isotope analyses of CV chondrite components (**Fig. 9**, see section 4.3; van Kooten et al., 2021; van Kooten and Moynier, 2019).

In summary, the MVE isotope systematics of carbonaceous chondrites are best explained by chondrule formation and subsequent combination with unmodified CI-like dust. The term 'unmodified' applies to the very limited amount of thermal processing this dust had to experience to maintain a CI-like MVE isotope composition. From there the question arises: is complementarity then just a matter of scale? How far did the matrix component in a chondrite have to have been from the chondrule reservoir to have preserved its MVE inventory?

### 4.2 Nucleosynthetic isotope variations

Mass independent variations are thought to reflect spatial and/or temporal variations in the isotopic composition of dust in the disk. Other than the gross distinction between the inner and outer Solar System, the scales at which these variations occurred have been the subject of much speculation but are still largely unknown. Nevertheless, mass independent isotopic variations can be used to infer mass transport between reservoirs. Mass independent isotope variations of bulk Solar System materials are typically small and require high precision analyses that necessitate the use of relatively large samples. The individual analysis of chondrules and matrix is size-limited and, as a result, has been mainly carried out on CV chondrites because their chondrules are large compared to other chondrite groups and CVs are relatively abundant. Hence, in understanding the direction and magnitude of mass transport of chondritic components in the disk through their nucleosynthetic isotope signatures, we have so far been restricted to a very small sampling of the Solar System that may well be overprinted by secondary processes (see section 5).

**Fig. 9: A)** *The $\varepsilon^{183}W$ values of bulk planetary bodies ($\varepsilon^{183}W = 0$) and the bulk, chondrule and matrix isotope compositions of Allende (from Budde et al., 2016),* **B)** *The $\varepsilon^{54}Cr$ versus $\delta^{66}Zn$ isotope composition of CV chondrule cores (n =5 chondrules, with large dark green sphere as average), their corresponding igneous rims, the bulk CV value and a fine-grained dust rim (FGR) surrounding one of the chondrules (modified after van Kooten et al., 2021). The bulk CV chondrite composition can be explained by mixing between inner Solar System derived chondrules that underwent progressive accretion of CI-like dust. The black mixing line shows 10% increments between mixing the chondrule cores and CI-like dust. The bulk CV plots at approximately 40% matrix, in agreement with estimates from Alexander (2019).*

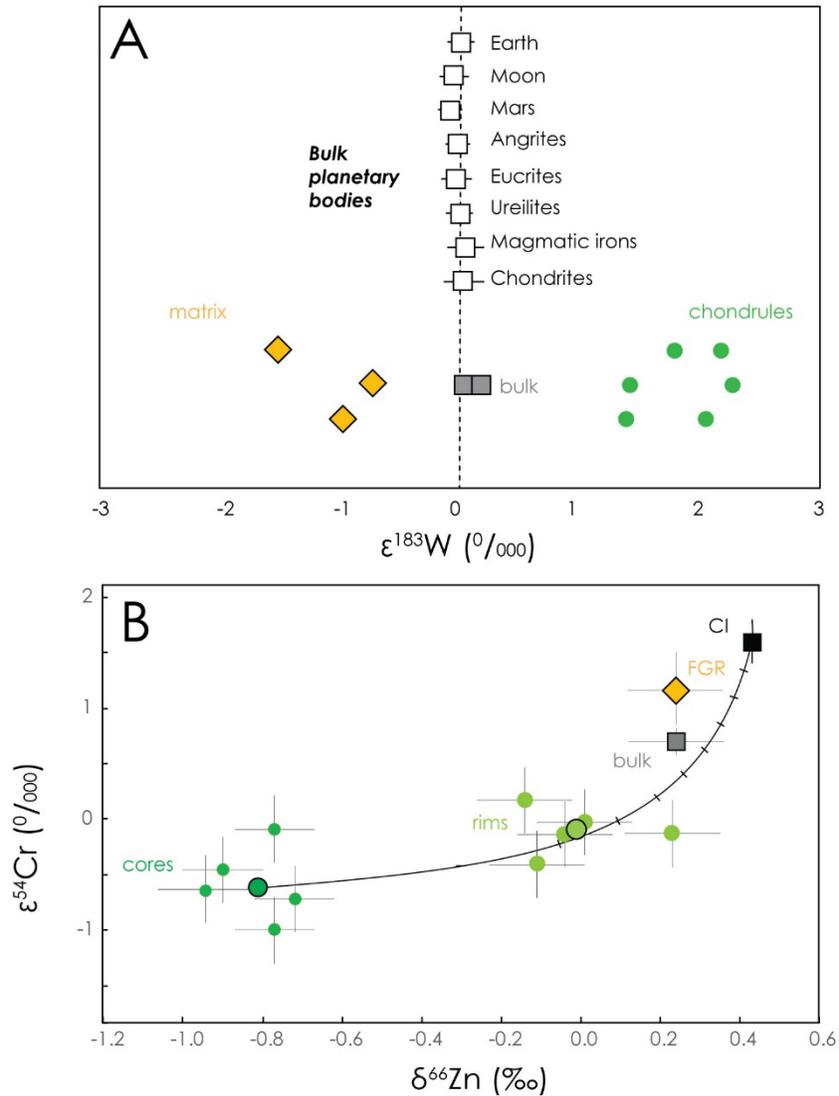

The nucleosynthetic W isotope systematics of Allende CV chondrules and matrix have so far been the greatest isotopic evidence for complementarity (**Fig. 9A**). While bulk Solar System materials, including CV chondrites, exhibit very limited variation in $\mu^{183}W$, the chondrule and matrix fractions of Allende are enriched and depleted relative to the bulk, respectively (Budde et al., 2016). As discussed above, a requirement of nucleosynthetic complementarity should be that matrix and chondrules have *identical* $\mu^{183}W$ signatures, since they form from the same presolar dust reservoir. However, Budde et al. (2016) have argued that 1) within the same reservoir, size sorting can occur that can preferentially enrich the chondrules in $^{183}W$-rich presolar grains (i.e., metal, sulfide or SiC), or 2) chondrules do not completely melt and then expel the presolar grains somehow. It is unclear how these models of chondrule formation would work in nature since the size sorting would require a larger scale over which complementarity operated to the extent that we must wonder if it can still be considered as such. For model 2, it begs the question why and in which relative abundance (to isotopically 'normal' grains) these relict presolar grains were expelled. Notably, CR chondrites contain significantly larger bulk $\mu^{183}$W values (Budde et al., 2018). This raises the question whether this isotope system emulates complementarity, or if it advocates for a general chondrule formation process, where the $\mu^{183}$W signature of all chondrules is supersolar and all matrix is subsolar and together forms a random isotopic mixture. An alternative explanation that has been put forward to explain the mirroring W isotope signatures is through secondary alteration (see section 5).

As the key tracers of (nucleosynthetic) isotope variability in the Solar System, Cr, Ti, and O isotopes are frequently used in combination to pinpoint mainly spatial differences in the accretion of (a)chondrite parent bodies (Trinquier et al., 2009; Warren, 2011). Since O and Ti isotopes each have their own complications with regards to the interpretation of the nature of chondrule and matrix precursor materials (i.e., secondary alteration, O precursors and gas interaction, refractory inclusions), we will continue discussing primarily the nucleosynthetic Cr isotope variations. Chromium isotope analyses have been carried out on individual chondrules from CV and CR chondrites (Olsen et al., 2016). The observed range of $\mu^{54}$Cr values is very restricted and positive for CR chondrules, whereas CV chondrules are highly variable and essentially cover the entire observed range of the Solar System materials. This has been interpreted as formation of CR chondrules in the outer Solar System (van Kooten et al., 2016) and formation of CV chondrules throughout the disk and subsequent mass transport to their final accretion region (Olsen et al., 2016) or as within a highly heterogeneous dust reservoir at their accretion location (Gerber et al., 2017; Williams et al., 2020; Schneider et al., 2020). Some of the nucleosynthetic Cr isotope variability in CV chondrules has also been attributed to AOA addition to the chondrules (Schneider et al., 2020). However, Hellman et al. (2020) have shown that the mass-dependent Te versus nucleosynthetic Cr isotope composition of bulk chondrites is best explained by a two-component mixing model between CI-like matrix and $^{54}$Cr-depleted chondrules. Recent analyses of individual CV chondrules from Leoville and their surrounding igneous and fine-grained accretion rims have shown the importance of Cr mobility during thermal metamorphism (see section 5; van Kooten et al., 2021). These analyses show that relatively unaltered CV chondrules display negative $\mu^{54}$Cr values, like ordinary chondrite chondrules, but become progressively enriched in $^{54}$Cr towards their outer rims (**Fig. 9B**). Note that Leoville chondrules, although significantly less altered than, for example, Allende, could still have experienced some Cr redistribution and these values should be considered as upper limits. As has been verified by Nie et al. (2021) through MVE isotope analyses (**Fig. 8**), this points to a two-component mixing between $^{54}$Cr-rich CI-like dust and $^{54}$Cr-poor chondrules. This has been interpreted as requiring significant mass transport of either chondrules to the outer disk or CI-like dust to the inner disk (Schiller et al., 2020, 2018; van Kooten et al., 2021) and does not support the complementarity model. Note that the Cr isotope matrix analyses only represent the fine-grained matrix surrounding the chondrules and not the coarser interchondrule matrix. The latter has a composition similar to the bulk CV chondrite and it is suggested to reflect matrix plus chondrule and refractory fragments that were processed on the parent body (van Kooten et al., 2021).

Collectively, with chondrule formation being a complex and poorly understood process, it is not straightforward to interpret the mass-dependent and nucleosynthetic isotope signatures of chondrules in the framework of complementarity. Working towards a solid understanding of isotopic complementarity requires 1) using unaltered samples, 2) chemical data of chondrules that allows us to assess the degree of elemental/phase loss from the chondrule relative to CI chondrites, 3) correlated mass-dependent and mass-bias corrected isotope data of chondrules and matrix, 4) isotopic analyses of chondrite groups other than CVs.

### 4.3 Radiometric isotope anomalies

The decay of long- and short-lived radionuclides has been used in cosmochemistry to obtain absolute (i.e., U-corrected Pb-Pb ages) and relative (i.e., $^{26}$Al-$^{26}$Mg, $^{53}$Mn-$^{53}$Cr, $^{182}$Hf-$^{182}$W etc.) ages of chondritic components. For a review about the caveats using each of these dating systems, the reader is referred to other review chapters (Connelly et al., 2017; McSween and Huss, 2022). Here, we briefly discuss the Pb-Pb and Al-Mg ages obtained for chondrules so far, assuming that these ages are accurate, and their implications for the chondrule-matrix complementarity debate.

From the perspective of U-corrected Pb-Pb ages obtained for individual chondrules from CV and CR chondrites, chondrules formed in a timeframe encompassing the entire disk lifetime of approximately 4 Myr (Bollard et al., 2019, 2017; Connelly et al., 2012). For CR chondrites, an offset between the U-corrected Pb-Pb ages of chondrules and the modeled accretion age of the CR parent body is observed. Neglecting the poor statistics of individual chondrule Pb-Pb ages for a moment, the mean CR chondrule formation age lies around 2.5 Myr after CAI formation (Bollard et al., 2019), whereas the accretion age is proposed to be around 4 Myr (Doyle et al., 2015; Schrader et al., 2015; Budde et al., 2018). If we assume that both ages are correct, this implies that CR chondrules were stored in the protoplanetary disk for >1 Myr, a timeframe that would have allowed a significant mass flux of fresh, 'anomalous' dust and pebbles to the CR accretion region. As such, the Pb-Pb ages of CR chondrules do not favor a complementarity model. For the CV chondrites, very few individually Pb-Pb dated chondrules exist (Bollard et al., 2019), and all have ages that are within 1 Myr after CAI formation. This agrees with pooled ages of Allende chondrules (4566.6±1.0 Ma; Amelin and Krot, 2007). The accretion age of CV chondrites is estimated to have been between 2-3 Myr after CAI formation (Doyle et al., 2015), suggesting again a gap between chondrule formation and accretion that allows for significant mass transport to the CV chondrite accretion region during that time.

In contrast to their Pb-Pb ages, $^{26}$Al-$^{26}$Mg ages of chondrules from the most primitive chondrites exhibit peak distributions close or coinciding with their accretion ages. For example, the observed peak of CC chondrules lies between 2-2.5 Myr after CAI formation (Pape et al., 2019; Villeneuve et al., 2009) From this perspective, storage of chondrules before accretion is not required and complementarity is more likely to be achieved in these circumstances. The reported Al-Mg ages of CR chondrules (3.5-4.0 Myr after CAI formation; Schrader et al., 2017) also show overlap with their accretion ages. However, Tenner et al. (2019) have reported that FeO-rich and FeO-poor CR chondrules formed at different time intervals, with a >1 Myr gap, again allowing for the possibility of mass transport during that time. Hence, resolving the age debate regarding the validity of either Pb-Pb or Al-Mg ages would go a long way in tightening the constraints around the complementarity debate.

## 5. Parent body alteration: the possibility of apparent complementarity

A major issue in understanding early disk evolution including dust processing, chondrule formation and mass transport, is separating these primary processes from subsequent secondary alteration in chondrite parent bodies. Establishing potential genetic links between chondrules and matrix requires the analyses of relatively unaltered samples and insights into the earliest stages of alteration, including the mobility of elements under various alteration conditions as well as their sources and sinks within chondrites. Studies investigating complementarity have so far focused dominantly on relatively altered chondrites, such as Allende (CV3.6 or higher), Murchison (CM2.5), Jbilet Winselwan (a shock heated CM2.1-2.4) and Renazzo (CR2.5). These chondrites have already experienced mild or more extensive thermal metamorphism and/or aqueous alteration, but it is not well understood how this affects each element that is the subject of a complementarity study. For example, some elements such as Cr, S, Fe, Ca, K, and Na are mobilized very early on, whereas other more refractory elements such as Al and Ti are typically less mobile. An assumption usually made about secondary alteration is that while the matrix is typically affected first (McSween, 1979; Bunch and Chang, 1980, Browning et al., 1996), it behaves as a closed system until relatively late in the alteration process when the matrix starts to interact with chondrules and other chondritic components at a local scale (Stracke et al., 2012). Hence, up to a certain stage (i.e., type 3 and 2 chondrites), even though the mineralogy of the matrix is completely altered, it is assumed we can use these chondrites to infer genetic relationships between *bulk* chondrules and matrix. Likewise, nucleosynthetic isotope compositions of chondrules and matrix are often assumed to reflect primary precursor dust signatures irrespective of the degree of secondary alteration. While this may be true for bulk compositions of chondrites, secondary alteration can result in mass-dependent isotope fractionation according to equilibrium laws or diffusion-related non-equilibrium isotope fractionation, processes that are not typically corrected for when reporting nucleosynthetic isotope data. In the following section, we discuss secondary alteration according to different element groups, the scale of alteration and the effect on the observed complementarity.

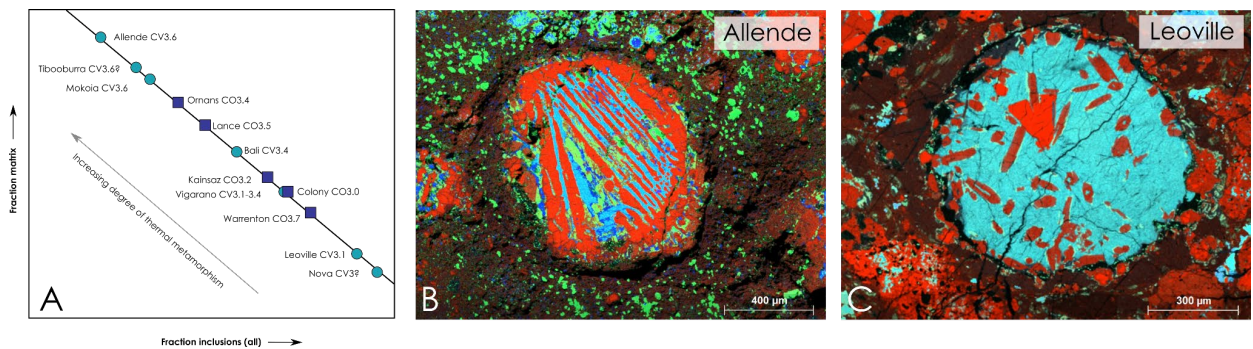

*Fig 10: Textural and compositional changes during progressive thermal metamorphism.  A) after Fig. 2 from (Ebel et al., 2016). In general, the fraction of the CV matrix increases with increasing petrological degree. Chondrules are eaten up and reprocessed into coarse-grained intra-chondrule matrices. CO chondrite fractions may be subject to a size sorting effect of increasing chondrule diameter with depth on the parent body and, hence, the petrological degree (Pinto et al., 2021). Similarly, the trend for CV chondrites has been suggested to reflect the pre-accretionary history of oxidized and reduced subgroups (Gattacceca et al., 2020). B) MgCaAl (RGB) elemental*

*map of Allende CV3.6 showing coarsening of the matrix, including Ca-rich pyroxene (green). The chondrule rims are being replaced with matrix-like material. **C**) MgCaAl map of Leoville CV3.1 showing an Al-rich chondrule in an apparently pristine state, including an outer rim of fine-grained matrix and coarser intra-chondrule matrix containing chondrule and CAI-like fragments.*

## 5.1 Siderophile elements (Fe, W, Mo)

Metal and sulfides are the main carriers of siderophile elements in chondrites. Therefore, the assessment of chemical complementarity between chondrules and matrix within this group of elements must consider the redistribution of these mineral phases during secondary alteration (and potentially during weathering). In the case of nucleosynthetic isotope complementarity, this assessment becomes more complex since small non-metal/sulfide phases (i.e., SiC, spinel and other presolar grains) may exist that together carry the anomalous components and behave differently from the metal and sulfide during secondary alteration.

As discussed above, siderophile elements and their isotopes are potential tools for testing complementarity (i.e., Fe contents, Hf-W contents, W and Mo isotope systematics). Since metal is one of the first phases to be altered during progressive oxidation by either fluid-assisted metamorphism or aqueous alteration, it is reasonable to assume that most, if not all, chondrites investigated for complementarity are affected by this process. If metal oxidation were to occur within a closed system represented by the matrix or the chondrules, this effect would be negligible. However, if siderophile elements were to transfer from chondrules to matrix and/or vice versa, this would pose a considerable problem to assess the degree of complementarity.

A complication in using siderophile elements is that metal grains in chondrules and their rims, as well as isolated larger grains in the matrix, are generally thought to be genetically related (Jacquet et al., 2013; van Kooten et al., 2022), where the larger matrix metal grains are suggested to be expelled from chondrules during their formation and reaccreted with the matrix. Hence, complementarity of siderophile elements is proposed to be achieved by physical redistribution of metal from chondrules to matrix when both components are present in the same reservoir (Hezel et al., 2018a). Alternatively, isolated metal grains that were roughly half the size of co-genetic chondrules (Skinner and Leenhouts, 1993) were transported along with chondrules to their final accretion region by aerodynamic sorting (Kuebler et al., 1999; Liffman, 2005), which does not require a genetic link between chondrules and matrix (van Kooten et al., 2022). In that sense, these larger metal grains represent metal-rich chondrules, in agreement with observations of igneous rims around the 'isolated' metal grains in the CM chondrite Paris (van Kooten et al., 2022). In contrast, the elemental composition of small metal grains (<50 μm) observed in the CM chondrite Maribo are significantly different from the larger metal grains, requiring an alternate origin (van Kooten et al., 2022). Furthermore, TEM images of Maribo matrix show abundant submicron-sized metal grains are present in the matrix, which all could be isotopically and chemically different from the larger chondrule-like metal grains (van Kooten et al., 2018). Hence, the composition of the unaltered fine-grained matrix may be considerably changed when chondrule metal grains start leaching siderophile elements into the matrix. These elements can be sourced from the isolated metal or from so-called armored chondrules that have metal rims connecting the chondrule with the matrix (van Kooten et al., 2019).

From a chemical perspective, oxidation and leaching of components in chondrule-derived metal and transport to the matrix would directly result in an increase of siderophile elements to the matrix, producing supersolar Fe/Mg and W/Hf ratios. In Mg versus Fe space, for relatively unaltered CV and CM matrices (i.e., Leoville and Maribo; Fig 1 in van Kooten et al., 2019), an increase in Fe relative to CI chondrites is observed, but no change in Mg content. This increase in Fe is suggested to be related to the first stage alteration of the matrix by oxidation of chondrule rim metal grains (i.e., armored chondrules) to the matrix. With increasingly altered CV and CM chondrites (i.e., Allende and Cold Bokkeveld, Fig. 1 in van Kooten et al., 2019), a correlated decrease in Fe and Mg in matrix is observed that is the result of exchange between chondrules and matrix (Hanowski and Brearley, 2001; Krot et al., 1995; Patzer et al., 2021, 2022, 2023). Hence, the unaltered matrix of CV and CM chondrites is suggested to have a CI-like Fe/Mg ratio. We note that this process has not been verified for other siderophiles, but caution should be taken when investigating complementarity with these elements. From an isotope perspective, it is unclear what the carrier phases of various W and Mo isotopes are and how they would behave with respect to each other during secondary alteration. It has been suggested that secondary chondrule-matrix exchange either involving selective dissolution of presolar carriers of s-process-depleted Mo and W in the matrix (Sanders and Scott, 2022), or oxidation of chondrule metal (Alexander, 2019) has resulted in the apparent isotope complementarity between chondrules and matrix in Allende. In CR chondrites (which notably have a distinct bulk $\mu^{183}$W from other Solar System materials), metal is found to be the most $^{183}$W-depleted phase (Budde et al., 2018). As such, the chondrite matrix would become increasingly $^{183}$W-depleted, relative to its starting composition, during secondary alteration. Finally, as evidenced by $\mu^{182}$W isochrons of Allende and Vigarano (a less altered CV chondrite), the matrix from Vigarano shows the same $\mu^{182}$W value as the bulk chondrite (Becker et al., 2015), whereas Allende has a subsolar $\mu^{182}$W for the matrix, suggesting that secondary alteration has caused the shift in $\mu^{182}$W.

## 5.2 Mg and Si

The subsolar Mg/Si ratio of carbonaceous chondrite matrices relative to the supersolar Mg/Si ratios of (mainly CV) chondrules is often used as a pillar of complementarity (see section 3). This assumes that the alteration of the matrix, which is observed to at least some degree even in the least altered chondrites, occurred in a closed system. However, loss of Si and Mg during progressive alteration of the matrix (and at later stage also from the chondrules; McSween, 1979; Bunch and Chang, 1980, Browning et al., 1996) is predicted for reactions that behave isovolumetrically, meaning that the volumes of the reaction products (e.g., serpentine) are equal to that of the reactants (olivine or pyroxene). Observations of pseudomorphic crystals of phyllosilicates replacing olivine and pyroxene in the matrix of CM chondrites are evidence of isovolumetric reactions. Therefore, the alteration from an anhydrous matrix to a hydrated serpentine-rich one is thought to proceed as follows (Velbel, 2014):

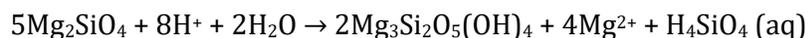

$$5Mg_2SiO_4 + 8H^+ + 2H_2O \rightarrow 2Mg_3Si_2O_5(OH)_4 + 4Mg^{2+} + H_4SiO_4 \text{ (aq)}$$

In this reaction, the resulting serpentine has a Mg:Si ratio of 3:2, whereas olivine has a Mg:Si ratio of 2:1. Hence, during isovolumetric serpentinization in an open system, the matrix is predicted to become enriched in Si over Mg. At later stages of aqueous alteration, the phyllosilicates become increasingly enriched in Mg over Fe, a process that is suggested to be facilitated by the alteration of olivine phenocrysts in chondrules. Hence, for a relatively unaltered chondrite matrix, we would expect to see subsolar Mg/Si ratios, whereas more altered matrices would exhibit Mg/Si ratios that approximate the bulk chondrite. This agrees with observations from CV, CM (Fig. S7 in van Kooten et al., 2019) and CO chondrites (Ebel et al., 2016). In addition, recent data from relatively unaltered CO, CM and CR chondrite matrices show subsolar Mg/Si ratios (Patzer et al., 2023, 2022, 2021), although this has been attributed to ~10% forsterite loss prior to accretion by the authors. Nevertheless, such serpentinization reactions represent endmember approximations in complex systems of water-rock interactions of which the scale is unknown. Hence, serpentinization of the matrix has the potential to change its Mg/Si ratio but the magnitude and direction of this change is dependent on the environmental parameters of alteration.

### 5.3 Refractory lithophiles (Ca, Sr, Ti and Al)

The abundances of refractory lithophile elements, such as Ca, Sr, Al, and Ti, are high in CAIs. Refractory inclusions in carbonaceous chondrites are not rare (~0.1–3 vol.%; Krot, 2019) and, thus, can be important sources of these elements to supply the matrix during aqueous alteration and fluid-assisted thermal metamorphism. For example, observations from Allende CAIs show that Ca, Sr and, to a lesser extent, Al were removed from the inclusions and presumably gained by the surrounding matrix (Krot et al., 2021). Calcium, for instance, was redistributed into carbonates and Ca-pyroxenes in the CV matrix (Murakami and Ikeda, 1994). Importantly, this textural alteration of the matrix also poses problems in identifying and isolating the primary fine-grained matrix component. In addition to the mobilization of refractory lithophiles in CAIs, significant addition of more volatile elements such as Fe, Si, Na, K and Cl is reported. Titanium, like Al, is relatively immobile during progressive thermal metamorphism. While detailed petrological descriptions have been made of secondary alteration and mineral replacement in CAIs from carbonaceous (Hashimoto and Grossman, 1987; Hutcheon and Newton, 1981; Keller and Buseck, 1991; Krot et al., 2021, 1995) and even non-carbonaceous chondrite groups (Dunham et al., 2023), little is known about the bulk CAI elemental flux. In the framework of complementarity, the latter is important to constrain the role of CAIs as source and sink during secondary alteration. Finally, chondrule mesostasis can also be an important carrier of some refractory lithophiles (e.g., Al-rich chondrules) and is readily altered. Leaching of Al and Ca from chondrule mesostasis has been observed in even relatively unaltered chondrules (Hanowski and Brearley, 2001; Brearley, 2006; Brearley and Krot, 2013; Patzer et al., 2023). Collectively, this implies that a significant fraction of the variability observed in the Ca and Al contents of altered chondrules and matrix (i.e., Allende and Y-86751; Hezel et al., 2007, 2008) can be ascribed to redistribution (i.e., Ca precipitation in the matrix) and coarsening (i.e., analyses including Al-rich refractory fragments in the matrix). This agrees with Ca and Al chondrule and matrix analyses from the Efremovka CV3.1 chondrite, showing distributions around the solar Ca/Al ratio for both components (Hezel and Palme, 2008).

### 5.4 Moderately volatile elements (Cr, Zn)

Nucleosynthetic Cr isotope systematics are a reliable tracer of genetic heritage in bulk Solar System materials (Trinquier et al., 2009, 2007; Warren, 2011). This tracer has also been measured inbulk chondrule and matrix fractions of Allende (Kadlag et al., 2019), in individual chondrules of CV, CK, CM, CR and CH/CB chondrites (Olsen et al., 2016; van Kooten et al., 2021, 2020, 2016; Zhu et al., 2019;

Williams et al., 2020; Schneider et al., 2020) and on individual matrix areas in CV chondrites (van Kooten et al., 2021). It has been known for some time, however, that Cr is easily mobilized at the onset of thermal metamorphism (Grossman and Brearley, 2005) and the $Cr_2O_3$ contents of fayalitic olivine in type II chondrules can be used to compare the relative degree of alteration between chondrites of a given parent body (Davidson et al., 2019). Hence, only the most unaltered chondrite samples (petrological type <3.1) should be used for the analyses of individual chondritic components. Indeed, van Kooten et al. (2021) have demonstrated that nucleosynthetic Cr isotope signatures of chondrules homogenize with the matrix towards the bulk CV composition during progressive thermal metamorphism. Hence, while bulk matrix and chondrule fractions separated from Allende now have identical Cr isotope signatures (Kadlag et al., 2019), this need not imply a genetic relationship. In contrast, Leoville chondrules exhibit distinct Cr isotope compositions from their surrounding fine-grained rims (van Kooten et al., 2021), suggesting that the chondrules and matrix, at least in reduced CV chondrites, are not genetically related.

Similar observations have been made for mass-dependent Zn isotope signatures of Allende and Leoville chondrule and matrix fractions. Allende chondrules display a relatively large variability in the Zn isotope signatures of individual chondrules (~1 ‰; Pringle et al., 2017), whereas individual Leoville chondrules have more restricted Zn isotope values (van Kooten and Moynier, 2019). This can be attributed to increasing sulfurization during thermal metamorphism, where the Zn content of chondrules increases by precipitation of sulfides within the chondrules (Zn being a chalcophile element). Other chalcophiles may exhibit similar behavior (e.g., Te, Sn, Cd) and caution in selecting meteorites should be applied when studying these elements.

In sum, we have demonstrated that secondary alteration can have a non-negligible effect on creating apparent complementary relationships between chondrules and matrix for some elements. However, as of now, we have a very limited understanding of what happens for each element (chemically and isotopically) during progressive thermal and aqueous alteration. To objectively determine the chondrule-matrix relationships, it is key to know the source and sink, the mobility during various P-T-$fO_2$ conditions and open/closed system behavior of each element.

## 6. The chemical and physical implications of complementarity

### 6.1 A single reservoir process and chondrule formation

An important requirement of complementarity is that chondrules and matrix were processed co-genetically in the same dust reservoir. However, the final chemical outcome, the chondrule and matrix compositions observed, depends very much on the mechanism(s) of chondrule formation, the feedback from the surrounding gas and the spatial scales at which the heating event(s) occurred. For example, if we consider chondrules to reflect 'CI-like dust minus a (volatile) component', matrix should consist of a complementary 'CI-like dust plus a (volatile) component'. In this context, if matrix is thermally processed alongside chondrules, why would matrix remain CI-like and not gain this volatile component from chondrules, and how does matrix maintain its primitive nature (e.g., presolar grain abundance, organic composition, volatile-rich inventory)? If we consider the nature of unaltered chondritic matrix, it generally consists of amorphous (Fe-rich) silicates with embedded nanosulfides, complex organic matter, rare anomalous presolar grains and anhydrous silicates (see section 2). As noted earlier, although this amorphous silicate material has been suggested to have affinities to GEMS in IDPs (GEMS-like), there are several significant caveats in relating GEMS (from IDPs) to GEMS-like materials (from chondrites) through secondary alteration (Leroux et al., 2015; Ohtaki et al., 2021). An important question is then whether the amorphous silicate fraction of primitive matrix represents a disequilibrium condensation product from the solar nebula (Brearley, 1993; Greshake, 1997; Wasson, 2008) or if they are altered GEMS-like material that were inherited from the interstellar medium (Ishii et al., 2018), or a combination of both (i.e., thermal processing of interstellar GEMS by FU Orionis outbursts; Alexander et al., 2017). If we consider the first, these amorphous silicate materials were formed alongside chondrules in the same heating event, during what has been called 'chaotic' disequilibrium condensation (Rietmeijer and Nuth, 2013) where the condensation sequence and the resulting composition are difficult to predict but is generally thought to be consistent with GEMS-like materials in chondritic matrix (Leroux et al., 2015; Ohtaki et al., 2021). The complex organic matter and presolar grains are then suggested to be a late addition to this thermally processed matrix, but only represent a small fraction of the matrix (5-10%; Alexander et al., 2017). As such, GEMS-like amorphous silicates make up the bulk of the matrix, which has been overall characterized as CI-like (Alexander et al., 2019; Zanda et al., 2018; van Kooten et al., 2019), possibly with the loss of 10 % forsterite to explain the subsolar Mg/Si ratios (Alexander et al., 2019; Patzer et al., 2021, 2022, 2023). An important question is then whether these proposed condensation products would have solar compositions, or if they should exhibit some volatile loss. One requirement would be that the heating occurs in an environment where the solid condensates remained fine-grained enough to stay coupled to the gas during cooling (Alexander et al., 2005, 2017). This is quite different to what seems to have happened during chondrule formation where densities were high enough to stabilize quite large liquid droplets and not condense very fine-grained solids.

## 6.2 Astrophysical implications

If chondrules formed in specific "chondrule-forming regions" and were then mixed with matrix that formed elsewhere prior to chondrite accretion, there is no inherent reason they should be complementary in chemical composition. The struggle here is to reconcile the phenomenology of chondrites, our observations, with the dynamical models that might describe protoplanetary disk processes. Plate tectonic theory took 60 years to reconcile observed geological phenomena with geophysical models of mantle convection.

Whether or not there is a genetic relationship between chondrules and matrix has far-reaching astrophysical implications for how protoplanetary disks function and planetesimals accreted material. If chondrule-matrix complementary indeed exists, this places several important constraints on the theorized and modeled aspects of Solar System formation.

First, chondrules must be melted locally andin the same reservoir as matrix, ruling out various chondrule-formation models (Wood, 1996b), including planetesimal collisions (Johnson et al., 2015; Asphaug et al., 2011), impact jetting (Sanders and Scott, 2012; Johnson et al., 2015; Cashion et al., 2022) and near-Sun melting followed by transport (Shu et al., 1996; Brownlee, 2014). Note that the mechanism of impact jetting includes the ejection of relatively unheated protomatrix from the surface of a planetesimal and is, henceforth, not necessarily considered as a caveat to the complementarity model (Cashion et al., 2022). Secondly, other models may be able to melt chondrules in the presence of matrix, but to a point where all matrix is destroyed, including the low-temperature components commonly found in pristine meteorites. These models include bow-shocks generated by relatively large planet(esimal)s (Hood and Horanyi, 1993; Hood, 1998; Desch and Connolly, 2002; Morris and Desch, 2010; Morris and Boley, 2018), shock waves from clumpy accretion (Boss and Graham, 1993), gravitational instabilities (Wood, 1996a; Boss and Durisen, 2005) and current sheets produced by magnetorotational instability, which have not been shown to sufficiently localize heating (McNally et al., 2013; Hubbard and Ebel, 2018; Lebreuilly et al., 2023). Hence, the complementarity model either requires an extremely localized heating mechanism or the later addition of a small amount of unheated non-complementary matrix component consistent with the conclusion of Braukmüller et al. (2018). Finally, chondrites of each class must accrete from their individual reservoirs on timescales faster than mixing of components between different reservoirs (Alexander et al., 2008; Alexander and Ebel, 2012). Jones (2012) found this timescale problem "one of the biggest conundrums" in our understanding, if we consider the formation of CAIs 2-3 Myrs prior to chondrule formation based on Al-Mg dating. This "storage problem" is a challenge for disk dynamical models.

Formation of chondrules coexisting with matrix in a single reservoir appears to require non-shock, non-collisional astrophysical mechanisms for chondrule formation that have not yet been conceived or fully investigated (Hubbard and Ebel, 2018). To date, large scale disk models rarely address radiative transfer, local ionization, and/or vertical disk structure. Recent ALMA observations indicate strong segregation of protoplanetary disks into radial zones separated by "pressure bumps" and/or planets (Dullemond et. al, 2018). Such regions, if persistent (Jiang and Ormel, 2021), could support a vertical structure that would engender strongly different physical-chemical regimes between a low-density region above one scale height and the dense, cold midplane. Heating at high vertical disk heights, in low-density regions ionized by external photons could provide a sufficiently local mechanism for chondrule formation. However, some chondrules retain volatile elements that would be isotopically fractionated in a low solid density, high-temperature regime (Alexander et al., 2008; Ebel et al., 2018), a feature that is not observed for numerous isotope systems (see section 4).

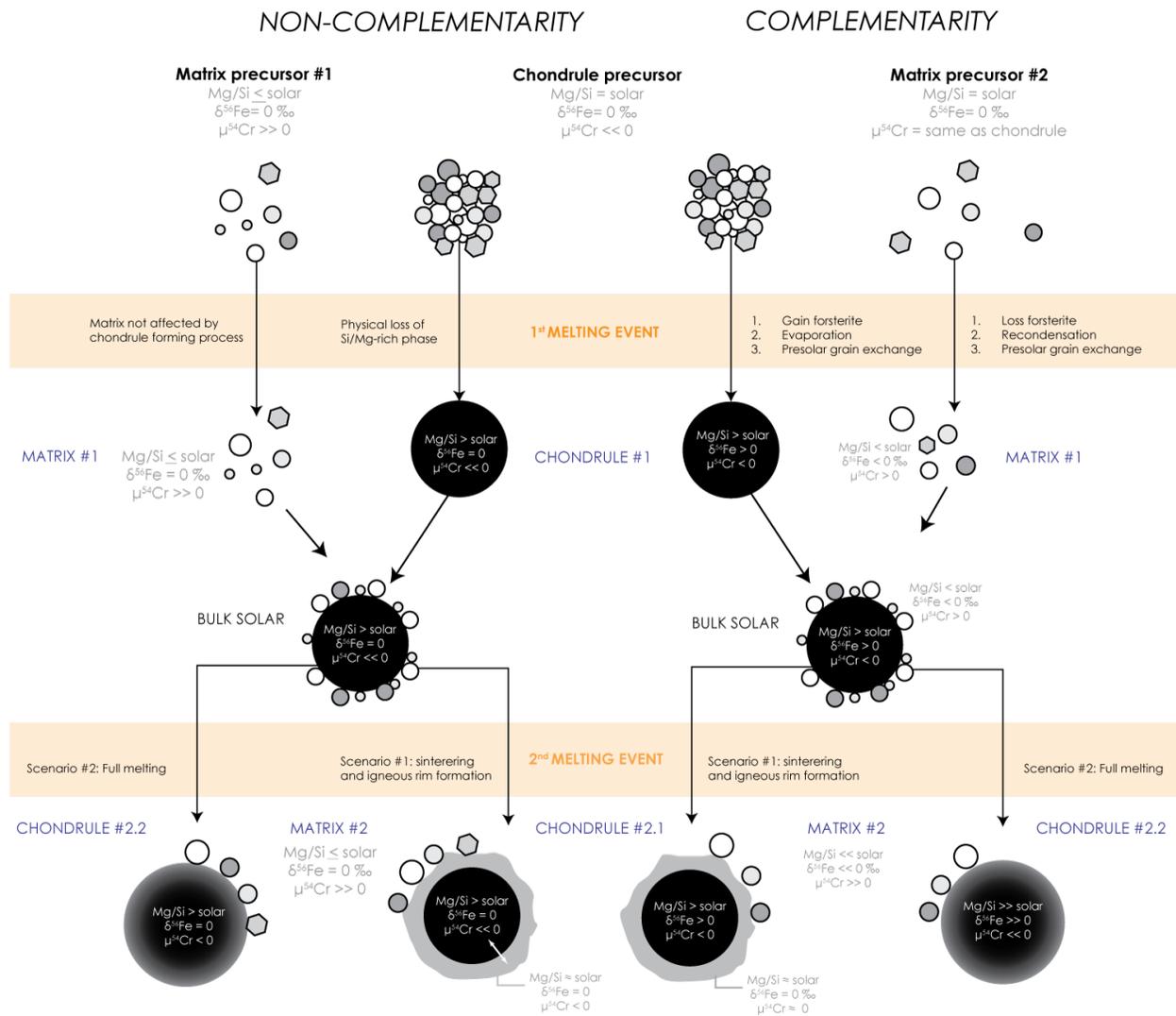

*Fig. 11: A schematic representation of chondrule formation and accretion in a complementarity and non-complementarity scenario and implications for the Mg/Si ratios, mass-dependent and mass-independent isotope signatures.*

## 7. How can we resolve the complementarity debate?

Our detailed discussion about the issues we face in resolving the complementarity debate, has left us with a few key follow-up points that we can address with future research:

- Understanding the formation pathways of GEMS-like matrix from either cold condensation the interstellar medium or 'hot' disequilibrium condensation in the protoplanetary disk will allow us to constrain whether chondrules and matrix formed independently or co-genetically, respectively. This requires expanding on the restricted amount of nanoscale petrological descriptions we have of highly primitive, unaltered matrices.

- In that respect, it is essential to restrict complementarity research to the most unaltered chondrites available to us, especially since we have a very limited understanding of elemental and isotopic redistribution and potential open system behavior during secondary alteration on the chondrite parent body and during terrestrial weathering. We should also consider that sample preparation can modify the composition of a sample. For instance, during polishing of samples, polishing compounds can be pressed into matrix, some phases may be soluble in lubricants/solvents that are used, others may be preferentially plucked because they are hard or removed because they are soft.
- A few key areas emerge on which we should focus our attention to correctly interpret the Mg/Si ratios of chondrules and matrix. First, although it appears that preliminary tests comparing EPMA with wet analyses of matrix cannot explain its subsolar Mg/Si ratio, further research is needed to systematically understand the effects of porosity and polyphase mineralogy that can offset EPMA from wet analyses data. Secondly, the broad definition of matrix used throughout the literature (grains < 5–100 μm) can include or exclude forsterite (but also pyroxene, metal and sulfide) grains that have a poorly defined origin. A deeper understanding of the relation of these grains to chondrules or matrix is critical to define the final Mg/Si ratios of these components. Finally, a detailed micron- to nano-scale investigation of these grains in pristine chondrites and their potential relationship to other primary matrix components can help constrain their abundance and origin and their importance in establishing subsolar Mg/Si ratios in the matrix of carbonaceous chondrites.
- Chondrule formation is a complex and poorly understood process, making it difficult to interpret the mass-dependent and nucleosynthetic isotope signatures of chondrules in the framework of complementarity. Improving on this issue requires 1) chemical data of chondrules that allows us to assess the degree of elemental/phase loss from the chondrule relative to CI chondrites, 2) correlated mass-dependent and mass-bias corrected isotope data of chondrules and matrix, 3) analyses of other chondrites groups than CV chondrites for isotope analyses.
- Constraining the relative timing and precursor material of multiple accretion events reflected by the onion-shell petrology of chondrules (i.e., igneous rims, primary and secondary chondrule cores) using consistent dating in combination with nucleosynthetic isotope systematics would significantly increase our understanding of the locality and frequency of chondrule formation and, therefore, on how we interpret the observed complementarity.
- Comparing cosmochemical datasets (i.e., chemical and isotope composition of chondrules and matrix versus chondrule/matrix abundance ratio and chondrule size distributions in chondrites) to dust and pebble growth and mass transport models that may apply in protoplanetary disks with real constraints from astronomical observations in the era of ALMA and JWST telescopes.

|  | Complementarity | Non-complementarity | Validation |
|---|---|---|---|
| The nature of matrix |  |  |  |
| GEMS-like material | IF: high T condensation origin | IF: cold condensation origin + alteration | Investigating primitive matrix on a nanoscale |
| Crystalline anhydrous silicates | H | x |  |
| Presolar grains | H | x |  |
| Organic matter | H | x |  |
| Ices |  | x |  |
| The nature of chondrules |  |  |  |
| Multiple melting events | ? | ? |  |
| Igneous rims |  | x |  |
| Chondrule populations | x | x |  |
| Chemical composition |  |  |  |
| Mg/Si ratio | IF: exchange of forsterite | IF: change in disk dust composition | Understanding the relationship of >5 μm matrix-embedded grains to matrix or |
| Volatility patterns | CR chondrites | x |  |
| Isotopes |  |  |  |
| Mass-dependent (Si, Mg, Fe) | ? | ? | Interconnection between chemistry, mass-dependent and nucleosynthetic isotope signatures of chondrules |
| Mass-dependent (MVE) | IF: exchange of MVE phases | x |  |
| Mass-independent | IF: exchange of presolar grains | x |  |
| Radiogenic | IF: Al-Mg ages are correct | IF: Pb-Pb ages are correct | High-prescion timeline of chondrule melting and accretion events |

**Table 2:** An overview of various constraints used in cosmochemistry to validate or disprove complementarity between chondrules and matrix. H = hybrid complementarity.

**Acknowledgements:** We thank two anonymous referees and editor Prof. Herbert Palme for their helpful comments that have been an improvement to this manuscript. The organizers of the International Space Science Institute workshop "Evolution of the Solar System: Constraints from Meteorites" (June 5-9, 2023, Bern) are thanked for their dedication to facilitate the writing of the review chapters.

**Conflict of interests:** The authors report no conflict of interests.

**Funding:**
Villum Young Investigator Grant (project no. 53024, EvK).